\documentclass[journal]{IEEEtran}

\usepackage[acronym]{glossaries}
\usepackage{acronym}
\usepackage{cite}
\usepackage{nameref}
\usepackage{amsmath}
\usepackage{amsfonts}
\usepackage{stackengine}
\usepackage{multirow}
\usepackage[table,xcdraw]{xcolor}
\usepackage{float}
\usepackage{placeins}
\usepackage{adjustbox, soul}
\usepackage{float}
\usepackage{graphicx}   
\usepackage{tabularx}
\usepackage{csquotes}
\usepackage{subcaption}
\usepackage{diagbox}
\usepackage{slashbox}
\usepackage{bbding}
\usepackage{epstopdf}
\usepackage{bm}
\usepackage{enumitem}
\setlist{nosep}

\usepackage{url}
\usepackage{booktabs}
\usepackage{hyperref}

\hypersetup{
    colorlinks=true,
    linkcolor=blue,
    filecolor=magenta,      
    urlcolor=cyan,
}

\newacro{avc}[H.264/MPEG-4 AVC]{}
\newacro{hevc}[H.265/HEVC]{}
\newacro{vvc}[H.266/VVC]{}
\newacro{sei}[SEI]{supplemental enhancement information}
\newacro{dct}[DCT]{discrete cosine transform}
\newacro{mhmcf}[MHMCF]{multi-hypothesis motion compensated filter}
\newacro{mrf}[MRF]{markov random field}
\newacro{ar}[AR]{autoregressive}
\newacro{cnn}[CNN]{convolutional neural network}
\newacro{bm3d}[BM3D]{block-matching and 3D filtering }
\newacro{dncnn}[DnCNN]{denoising convolutional neural networks}
\newacro{ffdnet}[FFDNet]{fast and flexible denoising convolutional neural
network}
\newacro{gan}[GAN]{generative adversarial network}
\newacro{cgan}[cGAN]{conditional generative adversarial network}
\newacro{bce}[BCE]{binary cross-entropy}
\newacro{msssim}[MS-SSIM]{multi scale structural similarity index}
\newacro{kadis}[KADIS-700k]{Konstanz artificially distorted image quality set}
\newacro{bce}[BCE]{binary cross-entropy}
\newacro{bsd}[BSD]{Berkeley segmentation dataset}
\newacro{psnr}[PSNR]{peak signal-to-noise ratio}
\newacro{ssim}[SSIM]{structural similarity index}
\newacro{jsdnss}[JSD-NSS]{Jensen Shannon divergence - natural scene statistics}
\newacro{mscn}[MSCN]{mean-subtracted contrast-normalized}
\newacro{jsd}[JSD]{Jensen Shannon divergence}
\newacro{nss}[NSS]{natural scene statistics}
\newacro{foe}[FOE]{fields of experts}
\newacro{hvs}[HVS]{human visual system}
\newacro{sei}[SEI]{supplemental enhancement information}
\newacro{jvet}[JVET]{joint video exploration team}
\newacro{mctf}[MCTF]{motion compensated temporal filter}
\newacro{mpeg}[MPEG]{motion picture experts group}
\newacro{mrf}[MRF]{Markov random field}
\newacro{vceg}[VCEG]{video coding experts group}
\newacro{ssd}[SSD]{Sum of Squared Differences}
\begin{document}

\title{Deep-based Film Grain Removal and Synthesis}

\author{{Zoubida~Ameur,
        Wassim~Hamidouche,
        Edouard~François,\\
     Miloš~Radosavljević,
         Daniel~Menard
         and Claire-Hélène Demarty}
\thanks{Zoubida Ameur, Edouard François, Miloš Radosavljević and Claire-Hélène Demarty are with InterDigital R\&D, Cesson-Sévigné, France (e-mail: firstname.lastname@interdigital.com).}
\thanks{Wassim Hamidouche and Daniel Menard are with Univ. Rennes, INSA Rennes, CNRS, IETR - UMR 6164, Rennes, France (e-mail: firstname.lastname@insa-rennes.fr).}}

\maketitle

\begin{abstract}
In this paper, deep learning-based techniques for film grain removal and synthesis that can be applied in video coding are proposed. Film grain is inherent in analog film content because of the physical process of capturing images and video on film. It can also be present in digital content where it is purposely added to reflect the era of analog film and to evoke certain emotions in the viewer or enhance the perceived quality. In the context of video coding, the random nature of film grain makes it both difficult to preserve and very expensive to compress. To better preserve it while compressing the content efficiently, film grain is removed and modeled before video encoding and then restored after video decoding. In this paper, a film grain removal model based on an encoder-decoder architecture and a film grain synthesis model based on a \ac{cgan} are proposed. Both models are trained on a large dataset of pairs of clean (grain-free) and grainy images. Quantitative and qualitative evaluations of the developed solutions were conducted and showed that the proposed film grain removal model is effective in filtering film grain at different intensity levels using two configurations: 1) a non-blind configuration where the film grain level of the grainy input is known and provided as input, 2) a blind configuration where the film grain level is unknown. As for the film grain synthesis task, the experimental results show that the proposed model is able to reproduce realistic film grain with a controllable intensity level specified as input.

\end{abstract}


\IEEEpeerreviewmaketitle
\begin{figure*}[t!]
\centering
  \includegraphics[width=\linewidth]{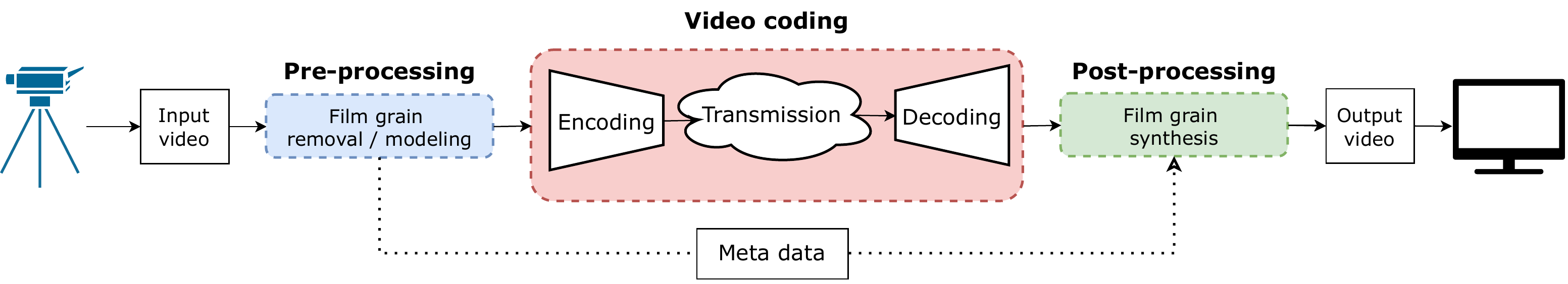}
  \caption{A simplified framework of the video distribution system with  film grain removal, modeling, and synthesis steps.}
  \label{fig:idea}
\end{figure*}

\section{Introduction}
Originally film grain is a characteristic of analog film. It is the result of the processes of exposing and developing silver halide crystals~\cite{salvaggio2013basic}, i.e., light-sensitive crystals that when exposed to light capture an image on a film. During the development process, crystals that are exposed to sufficient light are transformed into small particles of metallic silver. Others that are not developed are removed from film, leaving tiny gaps between those which are developed. Those small particles and gaps are in fact the result of many microscopic and chemical processes that, in the final stage of printing or projecting the film, lead to the creation of images with a grainy look.
Film grain appearance is therefore inevitable because of the physical nature of the process embedded in the film design itself. However, historically, it was considered as noise, and as such, technological advances have gone in the direction of its elimination.
With the arrival and evolution of digital camera sensors, film grain no longer exists. Moreover, digital imaging offered many more advantages in terms of robustness, reproducibility, and above all visual quality. Yet, most professional photographers and filmmakers would rather stick with the analog aspect when it comes to producing creative and artistic content, as they find the digital content too clean and sharp, which does not necessarily capture the sought atmosphere and therefore does not evoke the desired emotions. To better portray the cinematographic aspect of an analog film, several post-processing operations are commonly applied on the digital content. Adjustment of the color palette, adjustment of the contrast and generation of film grain contribute to distinctive characteristics of an analog film within a digital content \cite{newson2017stochastic}. This motivation turns film grain into a visual tool and not just a by-product of chemical processing as in analog film stock. 

However, within today's video distribution systems, the random nature of film grain makes its preservation difficult because of the high bitrate it requires to be encoded.  Therefore, it is challenging to find a balance between perfect fidelity of film grain and efficient compression that is an integral part of any such system \cite{norkin2018film,radosavljevic2021implementation}. Due to its random nature, film grain is difficult to predict by using typical prediction schemes of modern video coding standards. Thus, most of it remains in the prediction residue. Thereafter, its transformed coefficients are mainly distributed in the high frequency band, and are, consequently, more expensive to encode in the transform domain. The existence of film grain has a negative impact on the accuracy of predictions and motion estimation, which further reduces the coding efficiency in both motion estimation and spatial prediction~\cite{5495254}. Because of that, high bitrates are necessary to reconstitute the film grain with a sufficiently good fidelity \cite{radosavljevic2021implementation}. However, such high bitrates are generally not relevant in most common video applications.

To preserve film grain while improving coding efficiency, the natural approach would be to remove film grain from the content prior to encoding in order to achieve a higher coding efficiency and synthesize it back after decoding. When it comes to modern video distribution systems, where stronger compression is an inevitable step, film grain is destroyed at the encoder by compression itself without the possibility of reconstructing it. Hence, one solution is to use a parametric model to capture film grain characteristics prior to filtering and/or encoding and synthesizes it back at the decoder side with the aid of appropriate metadata. Figure \ref{fig:idea} provides a simplified block diagram of a typical video distribution system including film grain processing steps. Given an input video sequence, film grain is first filtered and modeled in a pre-processing step. The filtered video is then encoded and transmitted to the decoder together with film grain metadata. At the decoder side, the video is decoded and passed into a post-processing step that aims at reproducing the input video look by synthesizing film grain. Thus, film grain can be recovered while the content is more efficiently compressed. 

To summarize, film grain can be used: 1) in post-production where it is added to the digital content to improve its visual appearance and to add an artistic touch; 2) after decoding the content in case film grain was filtered during pre-processing and/or by compression itself (in which case model parameters are tuned manually or automatically to match or closely approximate the original look); or 3) and this is another potential use of film grain not already discussed above, as a visual tool tasked to mask compression artifacts and restore vividness in the compressed video (in which case it does not necessarily render the original film grain look and can be added even on content that had no film grain at the first place). In the latter case, film grain helps blending the content with its underlying texture, so that there is continuity between objects in the same image. It also helps smoothing out imperfections in the content, such as compression artifacts and distortions due to transmission errors~\cite{kurihara2011digital,wan2015improvement}.

Film grain is re-emerging in the age of digital content and is becoming increasingly relevant both for artistic motivations and perceptual quality enhancement. Concurrently, deep learning is nowadays applied in several computer vision and image processing tasks, with very impressive results due to the high modeling capability and advances in training and network design. To this end, we propose to leverage deep learning-based models to remove film grain before encoding and to synthesize it after decoding following the film grain encoding framework.

\section{Related work}

In this section, prior work related to film grain coding and processing is discussed. First, the film grain coding concept is defined as well as its standardization in different video codecs. Then, film grain coding stages are detailed, related first to film grain removal techniques as a pre-processing step, then to film grain synthesis techniques as a post-processing step. Film grain removal techniques are followed by an outline of the most interesting solutions in image denoising. Image denoising is one of the most fundamental tasks in image processing and computer vision, and de-graining is an equivalent task to denoising when film grain is considered as noise. The reader is referred to \cite{Tian2020DeepLO,goyal2020image} for a more detailed review on this topic. 
\label{sec:rel}
\vspace{-8mm}
\subsection{Film grain coding}
A typical modeling scheme for film grain coding was proposed by Gomila \textit{et al.} \cite{gomila2003sei}, in which film grain is filtered from the original video sequence as a pre-processing step before encoding and synthesized back as a post-processing step after decoding. A similar scheme has been used for speech coding \cite{benyassine1997itu}, where the inactive speech signal is pre-processed before encoding, and noise is added to the decoded signal for the comfort of human perception.
 
In the pre-processing step, film grain is further modeled and encoded in the form of a parameterized model. The model parameters are transmitted to the decoder and used to simulate back the film grain. The transmission of the parameters is accomplished by the so-called \ac{sei} messages. Since the introduction of a film grain \ac{sei} message in the H.264/MPEG-4 AVC \cite{gomila2003sei} standard, film grain modeling has become part of modern video coding standards. Following this, many works describing film grain parametrization and film grain synthesis within a video coding framework~\cite{sean_fghevc,radosavljevic2021grain,sean2021grain} were produced. For example, recent activity in the \ac{jvet} of ITU-T \ac{vceg} and ISO/IEC \ac{mpeg} promotes the modeling of film grain as part of the video distribution chain. The aforementioned works inherit the same syntax and semantics of the AVC film grain \ac{sei} message and apply it to the subsequent standards including H.265/HEVC and H.266/VVC. 
It is important to note that the \ac{sei} specification only provides the syntax to transmit parameters of the model, without the specification of the methodology for removing and synthesizing film grain or estimating model parameters. 
\vspace{-2mm}
\subsection{Film grain removal}
\label{subsec:fg_removal}
In the literature, algorithms specifically tailored for film grain removal were recently presented.
In \cite{schlockermann2003film}, it was proposed to use the  H.264/MPEG-4 AVC video encoder for film grain removal, where film grain is estimated by subtracting the encoded picture from the original one. Campisi \textit{et al.} \cite{campisi2000signal} proposed a filter that spatially adapts to local details in the image to avoid removing them while filtering film grain. The filter belongs to the class of contrast enhancement filters \cite{lee1980digital} and its coefficients are adaptively adjusted based on the local statistics of the image. Thus, removing film grain while avoiding adding distortions like blurring. A Bayesian approach can be incorporated into a physically motivated noise model where film grain is modeled using an inhomogenous $\beta$ distribution with the variance being a function of image luminance \cite{moldovan2006denoising}. To remove film grain, this model is combined with a recent prior model of images called \ac{foe} which is a high-order \ac{mrf} model that captures rich structural properties of natural images. 

A selective filtering using 2D spatial filters is applied only in the edge-free regions to remove film grain without blurring the edges and thus degrading the original image quality \cite{oh2007film}. This selective filtering preserves the edges and textures of the original image, but it somewhat limits the efficiency of the coding as film grain remains in the edges or the textured regions after denoising. In \cite{5495254}, prior to video encoding, essential parameters of film grain are estimated such as the spatial correlation of noise and the relationship between noise variance and signal intensity. Then, a temporal filter based on \ac{mhmcf} is applied to remove film grain. \ac{mhmcf} \cite{guo2006multihypothesis} is known for preserving most spatial details and edges, however, it was observed that film grain remains in the blue plane after applying the temporal denoising. Therefore, authors in \cite{hwang2013enhanced} proposed to explore cross-color correlations to enhance denoising performance.

The aforementioned work has investigated the film grain removal task and obtained qualitative results. However, most of the proposed solutions require at least one additional step before the removal of film grain, such as film grain modeling or edge detection. As a result, the quality of the filtered outputs is highly dependent on the quality of the outputs from the previous steps. Second, some methods use hand-crafted image features as prior information for filtering, which may not be relevant to the film grain removal task. Therefore, our solution is designed to overcome these drawbacks by using an end-to-end deep learning encoder-decoder architecture that takes a grainy image as input and outputs the corresponding filtered one. Since the input and the output are renderings of the same underlying content and structure, but with a different style, the low-level information is fed directly from the encoder to the decoder via skip connections such that the relevant underlying features for the film grain removal task are learned without any hand engineering. 
\vspace{-2mm}
\subsection{Image denoising}
Image denoising is a core image processing problem that has been studied for decades \cite{fan2019brief, goyal2020image, Tian2020DeepLO}. Many studies have tackled this problem and proposed different approaches \cite{dabov2007image,jain2013spatial,zhang2017beyond,tian2020attention,zhang2017learning}, each offering certain advantages and suffering from certain drawbacks making the image denoising task open and challenging. However, noise considered in most of these studies is assumed to be additive, non-correlated, Gaussian, and signal-independent, whereas film grain is signal-dependent and not necessarily Gaussian. Therefore, additional efforts had to be made to efficiently adapt state-of-the-art image denoising to the film grain removal task. 

Image denoising methods can be classified into two main categories: traditional hand-crafted and deep learning-based image denoisers. Traditional image denoisers can be roughly classified into two main sub-categories: spatial domain filtering and transform domain filtering. Denoisers that operate in the spatial domain are applied directly on the image samples to suppress the unwanted variations in sample intensity values and therefore, suppress noise. On the contrary, denoisers that operate in the transform domain first map the image into corresponding transform coefficients, on which they carry out some thresholding \cite{hou2007research}. As to learning-based image denoisers, state-of-the-art solutions are mainly based on \acp{cnn} which try to learn a mapping function by optimizing a loss function on a training set that contains pairs of clean references and noisy images \cite{zhang2017beyond, zhang2018ffdnet}.

\Ac{bm3d} is the most popular hand-crafted image denoising algorithm \cite{dabov2007image}. \ac{bm3d} is a non-locally collaborative filtering method in the transform domain. First, image patches similar to a given image patch are selected and grouped to form a 3D block, then, a 3D linear transform is performed on the 3D block followed by a filtering of the transform coefficients. Finally, an inverse 3D transform is applied to get back to the spatial domain and the different patches are aggregated to form the original image. This constitutes the first collaborative filtering step which is greatly improved by a second step using Wiener filtering. In the second step, instead of comparing the original patches, the filtered ones are compared and grouped to form the new 3D group which is processed by Wiener filtering instead of applying a threshold. Finally, an aggregation step is performed. A slightly different approach that utilizes temporal dimension of a video, named \ac{mctf} \cite{enhorn2020temporal, per2021temporal} is utilized within the latest \ac{vvc} reference software VTM. The proposed method relies on a bilateral filter \cite{710815} across  neighboring pictures compensated by  temporal motion. The \ac{mctf} is also used in \cite{radosavljevic2021grain} to filter out film grain from the video.

Several state-of-the-art works have addressed the problem of denoising using deep neural networks. Zhang \textit{et al.} \cite{zhang2017beyond} proposed a blind \ac{cnn}-based denoiser called \ac{dncnn} that takes as input a noisy image and outputs a denoised version of it. This work demonstrated that residual learning, originated in ResNet \cite{he2016deep}, and batch normalization, derived from Inception-v2 \cite{ioffe2015batch}, improve the denoising performance of the model. In a recent study, Zhang \textit{et al.} proposed a non-blind \ac{cnn}-based solution \cite{zhang2018ffdnet} \ac{ffdnet} that takes as input both the noisy image and its noise level map and outputs the denoised version. They proposed two different models for grayscale and color images of 15 and 12 convolution layers, respectively. The denoising is performed on downsampled sub-images to speed-up the training and to boost the performance. \Ac{dncnn} and \ac{ffdnet} have comparable architectures with a collection of Convolution + Batch Normalization + ReLU layers. Another model has been proposed by the same authors in their most recent work, DRUNet \cite{zhang2017learning}, where residual blocks were integrated into U-Net for effective denoiser prior modeling. Like \ac{ffdnet}, DRUNet can handle various noise levels via a single model. The experimental results shows that DRUNet achieves the best performance among all the state-of-the-art denoisers, both for grayscale and color images denoising.
\vspace{-2mm}
\subsection{Film grain synthesis}
In general, viewers tend to prefer images with a certain amount of fine texture such as film grain rather than sharp images \cite{kurihara2011digital ,wan2015improvement}. Since digital video is typically noiseless and since in many cases film grain is suppressed within various filtering and/or lossy compression steps, several studies have proposed film grain synthesis solutions.

In general, film grain synthesis approaches can be classified as signal-dependent or signal-independent. Signal-independent approaches involve applying a simple addition of or multiplication by a fixed and synthesized film grain to an image, where the synthesized film grain is either a stored example of film grain obtained by scanning and digitizing examples of film grain, or by the extraction of a grain pattern from real grainy images~\cite{schallauer2006film}. Signal-independent approaches are simple and fast. However, their results are deterministic, hence not suitable for \textit{synthesizing random} film grain. This results in a static film grain that can be very noticeable when applied to video sequences. However, film grain must be further blended according to the underlying signal in order to produce more realistic and pleasant visual appearance. One can classify signal-dependent film grain synthesis approaches into three main categories: mathematical-based models \cite{yan1997signal, yan1998film}, patch-based models \cite{efros1999texture} and parametric models based on texture statistics \cite{schallauer2006film}.

Mathematical-based models assume the presence of a pair of images, with and without film grain. In \cite{yan1997signal} and \cite{yan1998film}, higher order statistics are computed and used for noise parameter estimation and generation. However, the grain-free version of the image is not always known especially in real-world scenarios like streaming. In \cite{efros1999texture}, a patch-based model was proposed. It consists of a non-parametric method for sample-wise texture synthesis, where the texture synthesis process grows a new image outward from an initial seed, one sample at a time. To synthesize a single sample, first, regions in the sample image with small perceptual distance to the single sample’s neighborhood are gathered. The distance metric used to measure similarity between samples is the normalized \ac{ssd}. One of the regions is randomly selected and its center is used as the new synthesized sample in the context of an \ac{mrf}. As for parametric models based on texture statistics, in \cite{schallauer2006film} an adaptation of the parametric texture model approach \cite{portilla2000parametric} was adopted for film grain synthesis. First, the grain template image is decomposed into a steerable pyramid, a linear, multi-scale and multi-orientation image transform. Each scale and orientation of the pyramid are analyzed with respect to several statistical texture features including minimum and maximum gray values and correlation of sub-bands. The synthesis starts with random noise which ensures high spatio-temporal variations. The algorithm produces synthetic grain which matches the template very well while the random noise-based approach inherently provides superb spatial and temporal variations.

Based on two major and most advanced video coding standards, \ac{vvc} and AV1, film grain synthesis methods have also been experimented at the decoder side, using the \ac{vvc} and AV1 reference software implementations. For example, in the context of \ac{vvc}, to restore the film grain in the compressed video, a frequency filtering solution to parameterize and re-synthesize film grain can be used \cite{radosavljevic2021implementation,sean2021grain}. It is based on a low-pass filtering applied to the normalized Gaussian noise in the frequency domain. A film grain pattern is synthesized using a pair of cut-off frequencies, representing horizontal high cut-off frequency and vertical high cut-off frequency, which in turn characterize the film grain pattern (film grain look, shape, size, etc.). After the film grain pattern is obtained, it is scaled to the appropriate level using a stepwise scaling function which takes into account the characteristics of the underlying image. Afterwards, the film grain pattern is blended to the image by using additive blending. Likewise, in context of AV1, Norkin \textit{et al.} \cite{norkin2018film} propose to model the film grain pattern with an \ac{ar} model. Since film grain strength can vary with the underlying image intensity, they have proposed to reconstruct it by multiplying two terms, the film grain pattern generated by the \ac{ar} model and a piece-wise linear scaling function that scales film grain to the appropriate level before the result is added to the decoded image. 

Another autoregression approach is presented in \cite{oh2007film} where a 3D \ac{ar} model is used to model film grain considering the 2D spatial correlation and the 1D spectral correlation. Instead of scaling the generated film grain pattern as in the previous methods, the white signal used as a starting point for film grain generation is scaled. Similarly, in \cite{5495254}, film grain is modeled by an \ac{ar} model. Newson \textit{et al.} \cite{newson2017stochastic} proposed a stochastic model that approximates the physical reality of the film grain and designed a resolution-free rendering algorithm to simulate realistic film grain for any digital input image. This approach will be further detailed in subsection~\ref{dataset}.

Autoregressive models as well as frequency filtering-based methods enable the synthesis of a wide range of film grain patterns adapted to the content. In both cases, film grain is first analyzed and modeled at the encoder side with some parameters that are sent as metadata to the decoder for synthesis. The analysis stage requires a pair of samples with and without film grain and is performed only on the smooth / homogeneous regions because edges and texture can affect estimation of the film grain strength and pattern. Filtering and edge detection operations are performed for this. The synthesis step is also performed in two steps with film grain generated first and then scaled to the content using step- or piece-wise scaling functions. Considering this and driven by the ability of generative models \cite{10.5555/2969033.2969125} to generate realistic images, we propose to use a \ac{cgan} which thanks to its high modeling capacity learns a mapping function between grain-free and grainy images and thanks to the conditioning on the input image, the generated film grain is content-adapted. The \ac{cgan} choice is further motivated and inspired by the work in \cite{chen2020learning} where it was used for learning distortion generation. The main contributions of our work can be summarized as follows:
\begin{itemize}
    \item The first deep learning solutions for film grain removal and synthesis,
    \item Flexible deep learning-based film grain filtering and synthesis thanks to controllable intensity level,
    \item Content-adaptive and perceptually pleasant film grain synthesis,
\end{itemize}
The rest of this paper is organized as follows. Section~\ref{sec:rel} describes a brief overview on film grain removal and synthesis techniques in addition to state-of-the-art image denoising methods. Section~\ref{sec:proposal} provides a full description of the proposed solutions. In Section \ref{sec:experimental}, the experimental results are presented and analyzed. Finally, Section \ref{sec:conclusion} concludes the paper.

\begin{figure*}[t]
\centering
  \includegraphics[width=\linewidth]{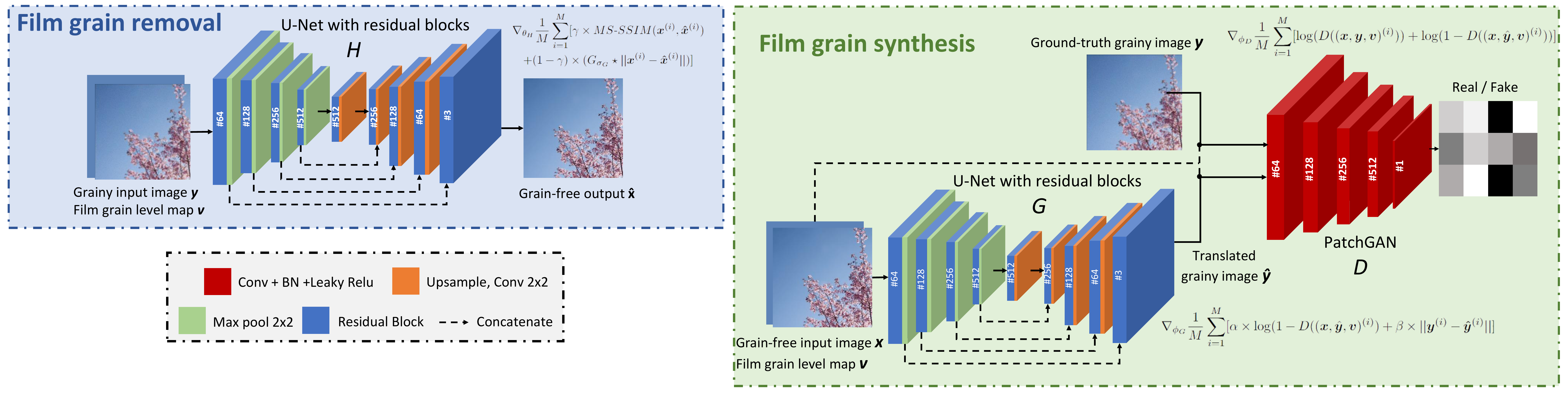}
  \caption{The framework of our proposed film grain removal and synthesis solutions.}
  \label{fig:endtoend}
\end{figure*}

\section{Proposed solution}
\label{sec:proposal}
In this paper, film grain removal and synthesis methods are proposed as described in Figure \ref{fig:idea}. Deep learning models and \acp{cnn} have proven to be very powerful and to outperform traditional techniques in several computer vision tasks, we propose to address each of these two steps using deep learning-based models for their ability to process and model large amounts of data. Network architectures as well as loss functions are chosen based on the properties and objectives of each task, i.e., removal or synthesis. 

\vspace{-2mm}
\subsection{Film grain synthesis}
\label{subsec:fg generation}
Film grain synthesis can be viewed as the translation of a given grain-free input image into a corresponding grainy output image while preserving the content. The goal is then to learn a mapping function from one input domain (grain-free images) $\bm{x}$ to another output domain (grainy images)  $\bm{y}$.
\begin{equation}
     \hat{\bm{y}} = G_\Phi ( \bm{x} ),
\end{equation}

where $G_\Phi$ is the parametric function of the film grain generation model and $\Phi$ its training parameters. 

A multitude of computer vision and image processing problems can also be modeled as image-to-image translation tasks including image synthesis \cite{reed2016generative}, image segmentation \cite{li2021semantic}, style transfer \cite{karras2019style}, etc. In \cite{isola2017image}, a \ac{cgan} was proposed as a general-purpose solution to image-to-image translation tasks motivated by the following two insights: 1) instead of hand-engineering a loss function to be minimized during training that satisfies the learning objective of each image-to-image translation task, \acp{cgan} learn automatically a loss adapted to the task and data at hand. 2) unlike \acp{gan}, \acp{cgan} learn the mapping by conditioning on an input and generating a corresponding output image. Since film grain is content-dependent and it is hard to manually design a loss function to be optimized for film grain synthesis, we propose to adopt a \ac{cgan} to solve the problem. Moreover, film grain synthesis has an artistic aspect, hence the use of a \ac{gan} where the goal is not to reproduce exactly the ground truth grainy images, but to generate realistic film grain while preserving the content.\\
\vspace{-2mm}
\subsubsection{Network architecture}
\hfill

Our proposed \ac{cgan} architecture is composed of a U-Net with residual blocks \cite{ronneberger2015u}, \cite{he2016deep} as generator and a PatchGAN as discriminator \cite{isola2017image}. The U-Net architecture was originally designed to tackle biomedical image segmentation. However, it has not only revolutionized medical imaging segmentation, but also other related areas such as image-to-image translation tasks \cite{zhang2017learning}. U-Net \cite{ronneberger2015u} is simply a U-shaped encoder-decoder with long skip connections between contraction and expansion levels, which represent its main feature. Residual blocks, on the other hand, were introduced as part of the ResNet architecture \cite{he2016deep}. Thanks to the local skip connections within each residual block, deeper networks with better performance were designed without the drawbacks of deep neural networks such as gradient vanishing and explosion. Combining the advantages of both U-Net and Residual blocks \cite{zhang2018road}, our proposed generator consists of a five scale U-Net model with residual blocks. Each residual block includes batch normalization, ReLU and convolutional layers. In addition, the skip-connections within the residual block consist in a convolutional layer to resize the output of the shortcut path to be of the same dimension as that of the main path. Unlike generator models in traditional \ac{gan} architecture, our proposed generator does not take a sample point from a latent space as input since it simply learns to ignore noise \cite{isola2017image}. 
To generate film grain at different intensity levels, the generator is conditioned by both a grain-free input image $\bm{x}$ and a film grain level map $\bm{v}$. 
\begin{equation}
     \hat{\bm{y}} = G_\Phi (\bm{x}, \bm{v}),
\end{equation}

The film grain level map $\bm{v}$ is a channel of the same dimensions as the input image, where all pixel values are equal to the film grain level of the corresponding target grainy image during training  such that $\bm{v}: R^{2} =  \{0.010, 0.025, 0.050, 0.075, 0.100 \}$.  Thus, a single model is used to generate film grain at different intensity levels by tuning only the film grain level map. 

The architecture of discriminator $\bm{D}$ is based on the PatchGAN architecture \cite{isola2017image} with a $30 \times 30$ receptive field. The discriminator takes as input two pairs of images: 1) The grain-free input and the grainy ground truth image with its corresponding film grain level map, which it should classify as genuine. 2) The grain-free input and the grainy translated image (output by the generator) with its corresponding film grain level map, which it should classify as fake. PatchGAN tries to classify whether each $70 \times 70$ patch in an image is fake or real instead of providing a single probability  for the entire input image. The use of PatchGAN limits the discriminator's attention to the local structure of the patch. Thus, it only penalizes the patch-scale structure and learns to model high frequencies. Similarly, the discriminator is trained to distinguish real ground truth grainy images from the ones translated by the generator conditioned by both grain-free input images and film grain level maps, such that it does not tolerate the generator to produce nearly the exact same output regardless of the input content nor the film grain intensity level. The detailed architecture of the proposed \ac{cgan} is illustrated in Figure~\ref{fig:endtoend}.
\subsubsection{Loss functions}
\hfill

During training, the generator aims to produce realistic grainy images, close to the ground truth ones, in order to fool the discriminator. Concurrently, the discriminator aims to correctly discern genuine grainy images from those translated by the generator. 
This leads the \ac{cgan} to model a conditional distribution of the target image $\bm{y}$ given both a grain-free input image $\bm{x}$ and a film grain level map $\bm{v}$, with the objective function $\mathcal{L}_{cGAN} (G, D)$ given by:
\begin{gather}\label{eq:gan}
    \mathcal{L}_{cGAN} (G,D) =  \mathbb E_{\bm{x},\bm{y},\bm{v}}[\log(D(\bm{x},\bm{y},\bm{v}))] + \\  \nonumber
    \mathbb E_{\bm{x}, \hat{\bm{y}}, \bm{v}} [\log (1 - D(\bm{x},\hat{\bm{y}} ,\bm{v})],
\end{gather}

where $G$ tries to minimize this objective against an adversarial $D$ that tries to maximize it, i.e., $G^{*} = \textrm{arg} \textrm{min}_{G} \textrm{max}_{D} \mathcal{L}_{cGAN} (G, D)$. 

Several approaches have shown that combining the \ac{cgan} objective with a more traditional loss, such as  $\ell_{1}$ or $\ell_{2}$ distance, yields better performance \cite{isola2017image, pathak2016context}. In most computer vision tasks, $\ell_{2}$ is the default loss function optimized during training. However, the latter penalizes high errors and tolerates small ones. Therefore, it assumes that the impact of noise is independent of the local characteristics of the image, while the \ac{hvs} is more sensitive to luminance, contrast, and structure~\cite{wang2004image}.  On the contrary, $\ell_{1}$ does not over-penalize larger errors and has proven to be more efficient when the task involves image quality \cite{zhao2016loss}. Accordingly, we combined the $\ell_{1}$ distance with the \ac{cgan} objective loss to optimize our model. The $\ell_{1}$ distance-based loss $\mathcal{L}_{L_{1}}$ represents the pixel difference between ground truth and translated images and is defined as:
\begin{equation}
    \mathcal{L}_{L_{1}}(G) = \mathbb E_{\bm{x}, \bm{y}, \bm{v}} \left [\lvert \lvert  \bm{y} - G(\bm{x},\bm{v})\rvert \rvert _{1} \right ] ,
\end{equation}

The generator $G$ is then trained to minimize a pixel-to-pixel error with the \ac{cgan} objective loss. The \ac{cgan} objective loss represents feedback from the discriminator and reflects whether the latter has been tricked or not. This feedback helps the generator learn the mapping of the ground truth image distribution, and thus, controls the perceptual quality. On the other hand, the discriminator $D$ is trained to distinguish between grainy ground truth and grainy translated images by maximizing the \ac{cgan} objective loss. This leads the final objective function to be defined as:
\begin{equation}
\label{eq:synthesis_loss}
    G^{*} = \textrm{arg} \textrm{min}_{G} \textrm{max}_{D} \mathcal{L}_{cGAN} (G, D) + \lambda \, \mathcal{L}_{L_{1}}(G),
\end{equation}

$\lambda$ is a weighting factor that controls the contribution of the $\mathcal{L}_{L_{1}}$ loss in the training process of $G$.
\vspace{-2mm}
\subsection{Film grain removal}
\label{subsec:fg removal}
The film grain removal task can as well be modeled as an image-to-image translation task where the goal is to learn a mapping from one input domain (grainy images) $\bm{y}$ to another output domain (grain-free images) $\bm{x}$. We propose, in this paper, two configurations of the same model, a blind version and a non-blind version. The blind version takes as input only the grainy image $\bm{y}$ while the non-blind version is provided with a grainy image $\bm{y}$ and its corresponding film grain level map $\bm{v}$ as input.

\begin{equation}
\label{eq:nonblind}
\begin{cases}
    \hat{\bm{x}}=  H_{\theta_1} (\bm{y}, \bm{v})& \text{non-blind}\\
     \hat{\bm{x}}=  H_{\theta_2} (\bm{y})& \text{blind}
\end{cases}     
\end{equation}

where $H_{\theta_1}$ is the parametric function of the non-blind film grain removal model and $\theta_1$ its training parameters and  $H_{\theta_2}$ is the parametric function of the blind film grain removal model and $\theta_2$ its training parameters. $\bm{v}$ is the corresponding film grain level map of the grainy input image $\bm{y}$.

Only the encoder-decoder architecture from the proposed \ac{cgan} is adopted to solve film grain filtering task but with different inputs and outputs as illustrated in Figure \ref{fig:endtoend}. 
\subsubsection{Loss functions}
\hfill

Unlike the film grain synthesis task, for which it is difficult to manually design a loss to be minimized during training, the film grain removal task can be learned by simply minimizing a pixel-to-pixel difference such as $\ell_{1}$ and/or a perceptual quality measure such as the \ac{ssim} \cite{wang2004image}. The film grain removal task consists in learning to properly filter film grain and restore as close as possible the grain-free ground truth image without introducing any additional distortions to the content such as: loss of detail, change in brightness or color shift. In order to fulfill all these requirements, we have opted for a weighted sum of a pixel-to-pixel loss $\mathcal{L}_{L_{1}}$ and a perceptual loss $\mathcal{L}_{MS-SSIM}$ as in \cite{zhao2016loss} for training our model. Authors in \cite{zhao2016loss} were the first to propose a mix loss function that combines $\ell_{1}$ and the \ac{msssim} \cite{wang2003multiscale} for training deep learning models to solve multiple image processing problems including denoising and demosaicking, super-resolution and blocking artifacts removal. In all tasks, it has been proved that \ac{msssim} helps preserve the contrast in high frequency regions while $\ell_{1}$ helps preserve color and luminance, therefore combining them provided relatively better results both in objective and subjective evaluations. \ac{msssim} is a full-reference quality metric that, for a given filtered image $\bm{\hat{x}}$ and a corresponding reference image $\bm{x}$, is defined as: 
\begin{gather}
        MS\textrm{-}SSIM(\bm{x}, \hat{\bm{x}})= l_{M}^{\alpha}(\bm{x}, \hat{\bm{x}}) \,  \prod_{j=1}^{M} cs_{j}^{\beta_{j}}(\bm{x}, \hat{\bm{x}}) \\ \nonumber
        l(\bm{x}, \hat{\bm{x}}) =  \frac{ 2 \mu_{\bm{x}} \mu_{\hat{\bm{x}}} + C_{1}}
    {\mu_{\bm{x}}^2 + \mu_{\hat{\bm{x}}}^2 + C_{1}} , 
     cs(\bm{x}, \hat{\bm{x}}) = \frac{2 \sigma_{\bm{x}\hat{\bm{x}}} + C_{2}} { \sigma_{\bm{x}}^2 +  \sigma_{\hat{\bm{x}}}^2 + C_{2}} 
\end{gather}
 
$\alpha$ and $\beta_{j}$ are parameters to define the relative importance of the components and are set to $\alpha = \beta_{j} = 1$ as in the original paper. $\mu_{x}$, $\mu_{\hat{x}}$ and $\sigma_{x}$, $\sigma_{\hat{x}}$ are means and variances of $x$ and $\hat{x}$ respectively. They can be viewed as estimates of the luminance and contrast of $x$ and $\hat{x}$, while $\sigma_{x\hat{x}}$ measures the tendency of $x$ and $\hat{x}$ to vary together, thus indicating structural similarity. 
$C1$ and $C2$ are used to stabilize the division and are defined as $C1 = (k_{1}L^{2})$, $C2 = (k_{2}L^{2})$ with $k_{1} = 0.01$, $k_{2}=0.03$ by default and $L$ being the dynamic range of the pixel-values. $M$ represents the scale number at which \ac{msssim} is computed.

Optimizing a model using \ac{msssim} as loss function consists in maximizing the latter which is equivalent to minimizing the following equation: 
\begin{equation}
\mathcal{L}_{MS-SSIM} = 1 - MS\textrm{-}SSIM(\bm{x}, \hat{\bm{x}})
\end{equation}

Moreover, since $\mathcal{L}_{MS-SSIM}$ propagates error at a given pixel based on its contribution to $MS\textrm{-}SSIM$ of the central pixel according to the filter size, $\mathcal{L}_{L_{1}}$ is weighted by the same Gaussian filter $G_{\sigma_{G}}$ of size $11\times11$ and variance $\sigma_{G}=1.5$ used in $MS\textrm{-}SSIM$. Therefore, our film grain removal model $H$ is optimized by minimizing the following mix loss function:
\begin{gather}
\label{eq:filtering_loss}
     \mathcal{L}_{H} = \gamma  \, \mathcal{L}_{MS\textrm{-}SSIM} + ( 1 -  \gamma) \, (G_{\sigma_{G}}  \star \mathcal{L}_{L_{1}} )
\end{gather}
where $\gamma$ represents a weighting factor for loss functions and is set to 0.84 and $\star$ refers to the convolution operation. 

\begin{figure}[b!]
\centering
    \begin{subfigure}[]{\linewidth}
        \includegraphics[width=0.3\linewidth]{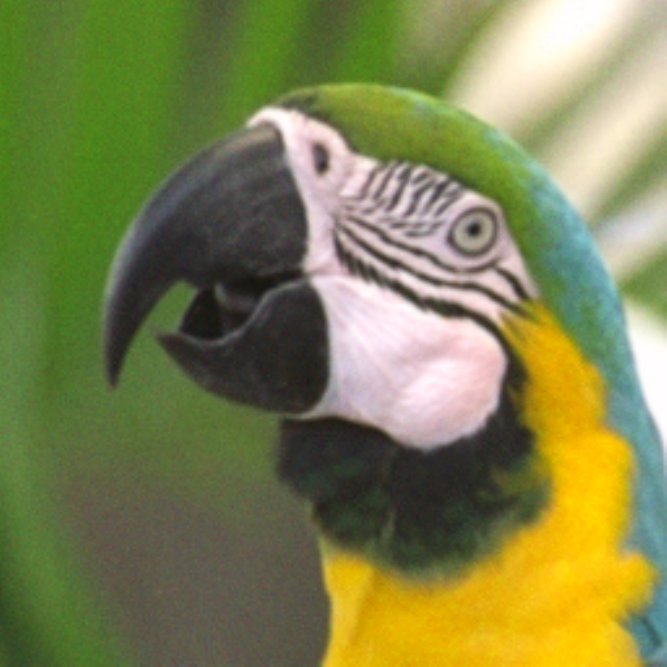}\hfill
        \includegraphics[width=0.3\linewidth]{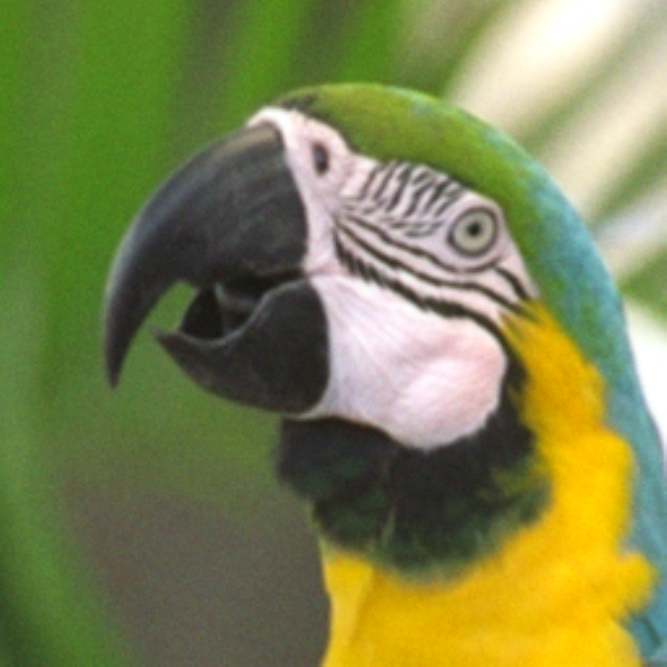}\hfill
        \includegraphics[width=0.39\linewidth]{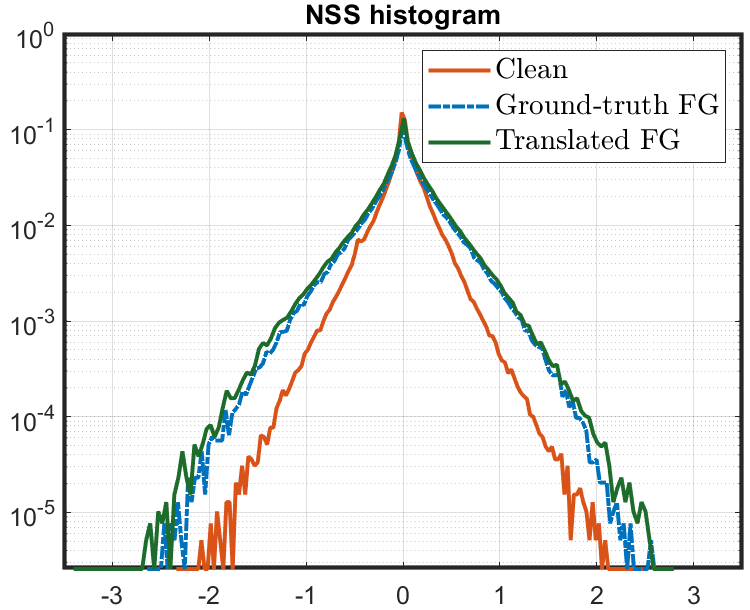}
        \caption{FG level = 0.010 ($1.64\mathrm{e}{-4}$)}
    \end{subfigure}
    
    \vspace{.6ex}
      \begin{subfigure}[]{\linewidth}
    \includegraphics[width=0.3\linewidth]{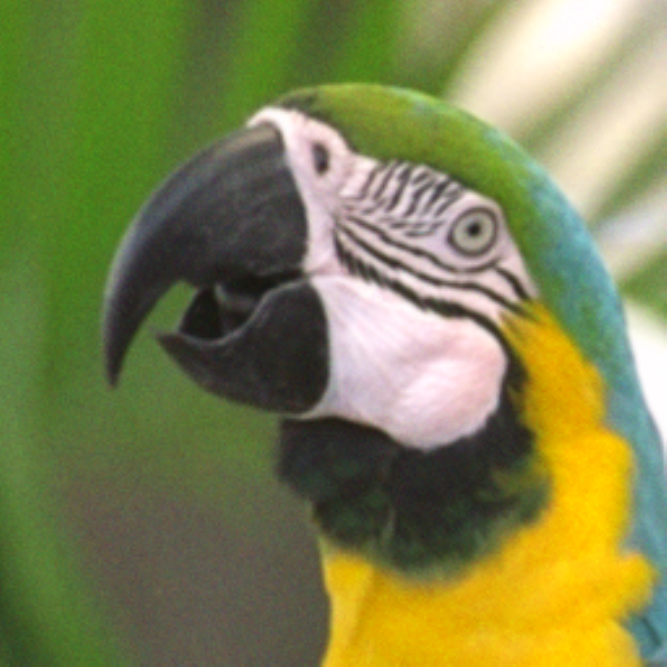}\hfill
    \includegraphics[width=0.3\linewidth]{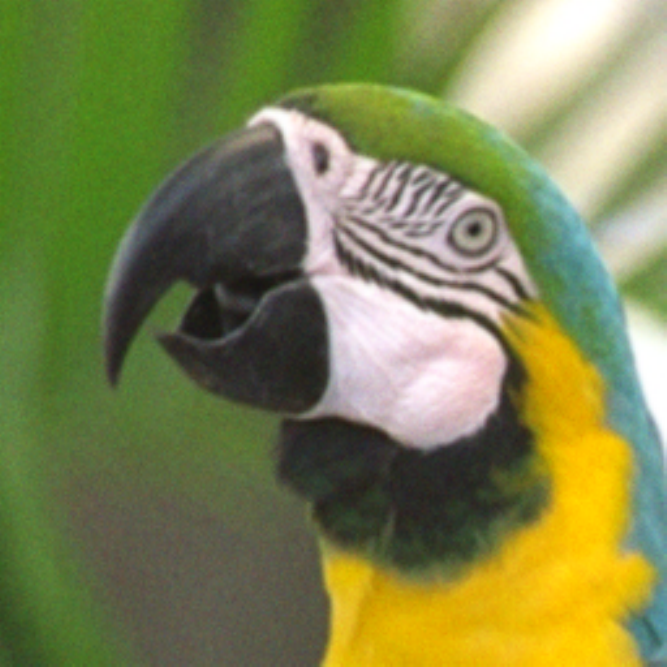}\hfill 
    \includegraphics[width=0.39\linewidth]{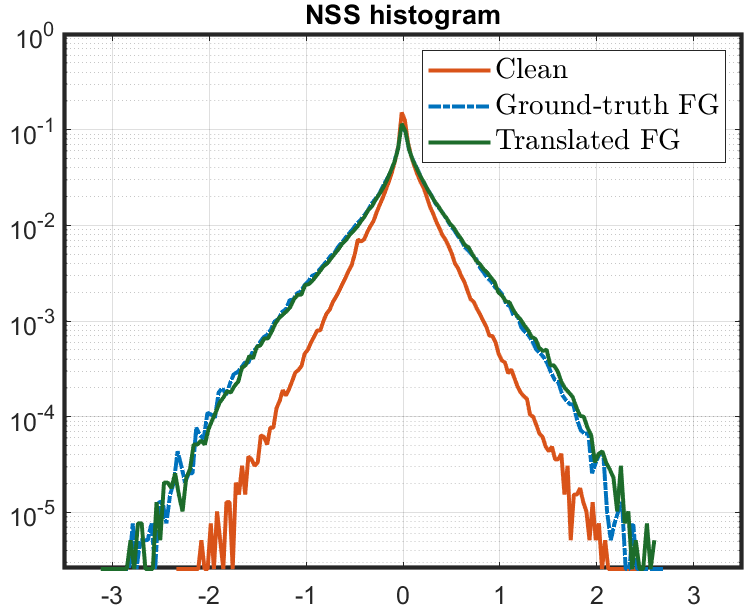}
    \caption{FG level = 0.025 ($1.61\mathrm{e}{-4}$)}
    \end{subfigure}

    \vspace{.6ex}
      \begin{subfigure}[]{\linewidth}
    \includegraphics[width=0.3\linewidth]{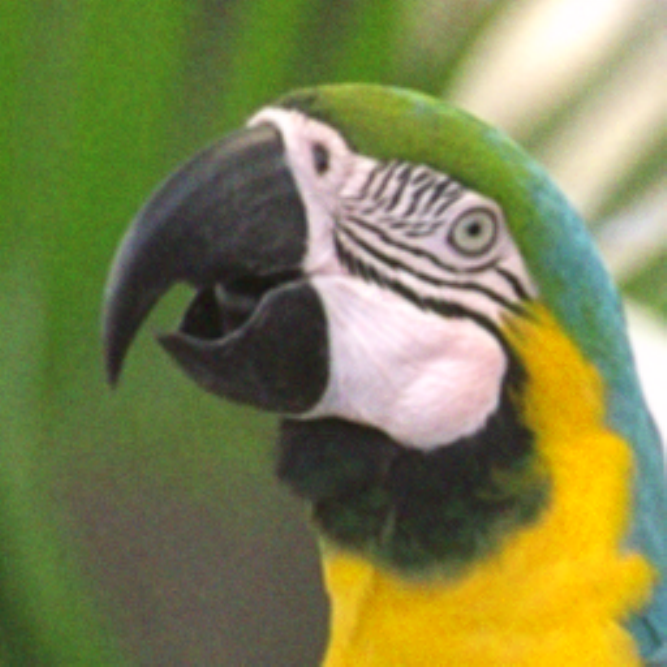}\hfill
    \includegraphics[width=0.3\linewidth]{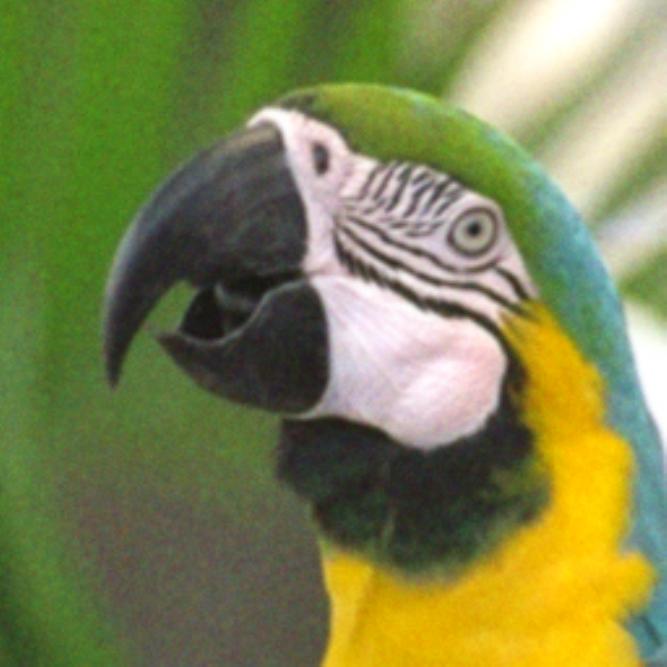}\hfill
    \includegraphics[width=0.39\linewidth]{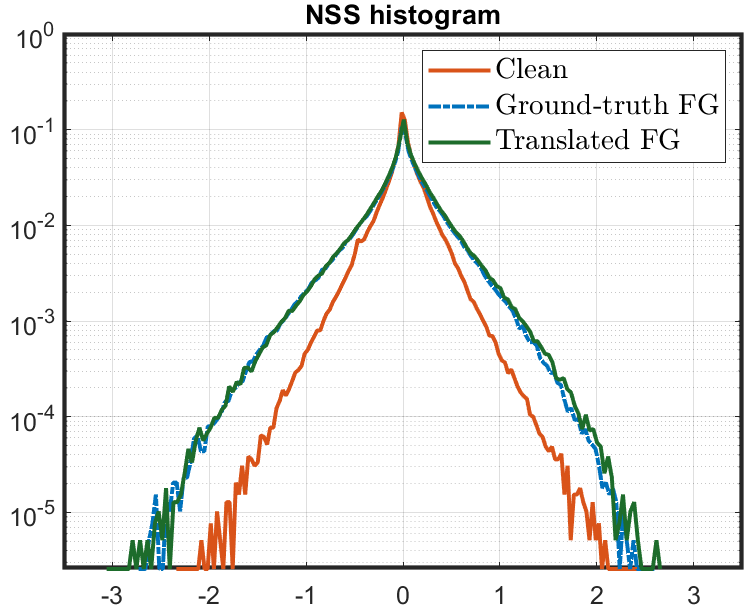}
    \caption{FG level = 0.050 ($1.29\mathrm{e}{-4}$)}
    \end{subfigure}    
    
    \vspace{.6ex}
      \begin{subfigure}[]{\linewidth}
    \includegraphics[width=0.3\linewidth]{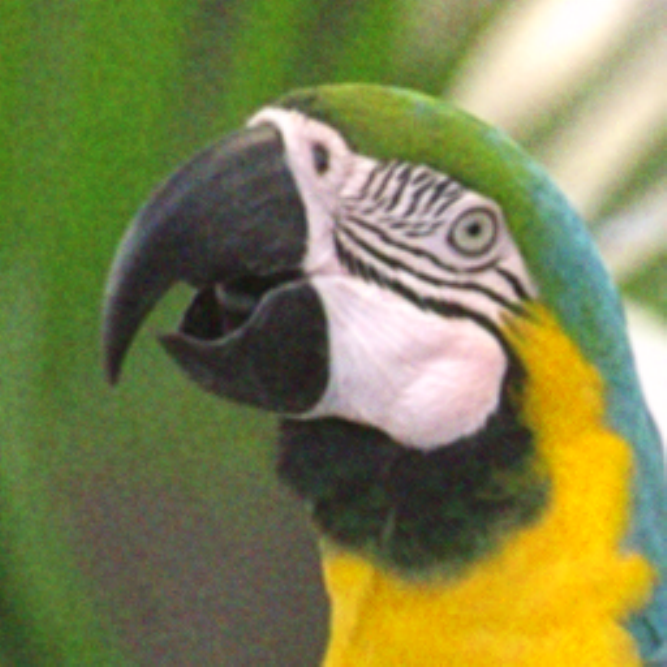}\hfill
    \includegraphics[width=0.3\linewidth]{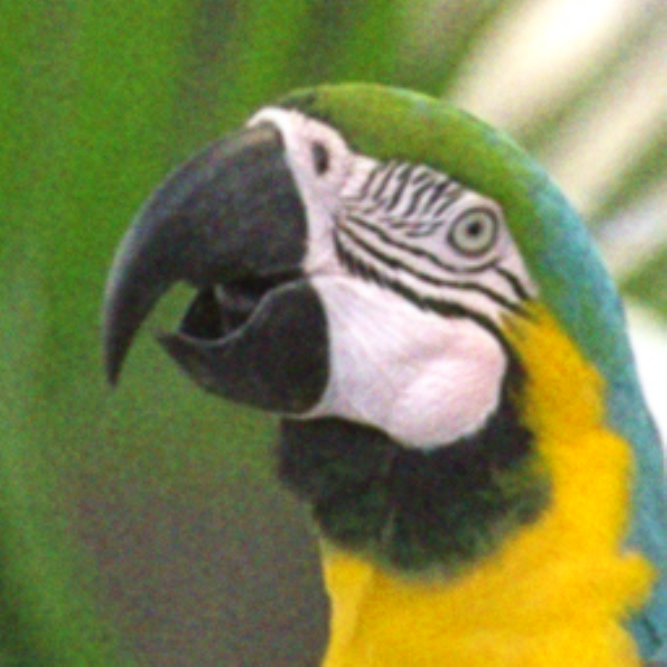}\hfill
    \includegraphics[width=0.39\linewidth]{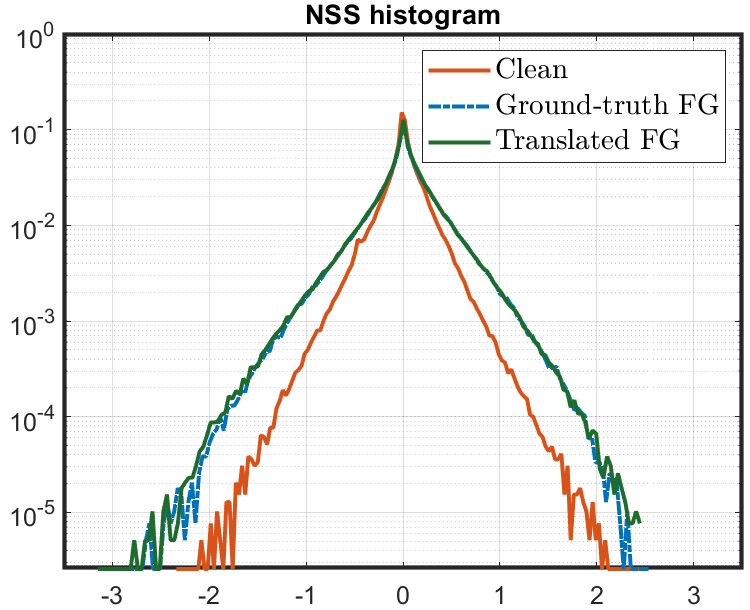}
    \caption{FG level = 0.075 ($8.42\mathrm{e}{-5})$}
    \end{subfigure}      
    
    \vspace{.6ex}
      \begin{subfigure}[]{\linewidth}
    \includegraphics[width=0.3\linewidth]{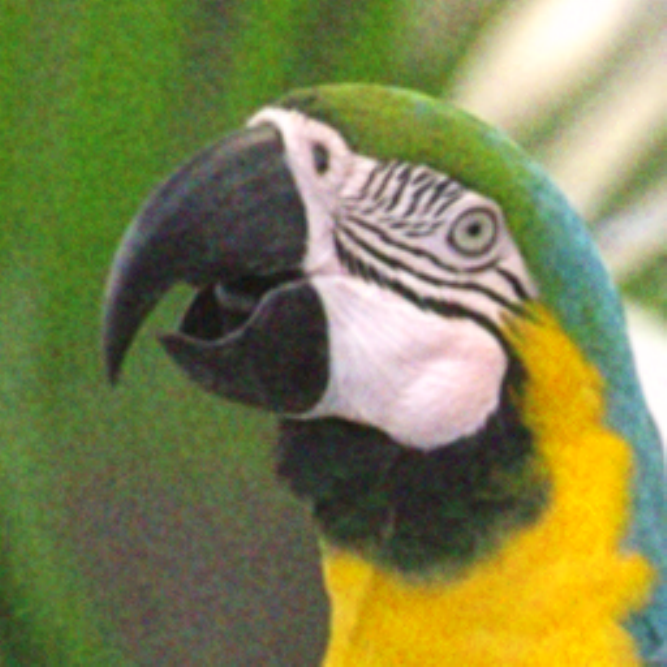}\hfill
    \includegraphics[width=0.3\linewidth]{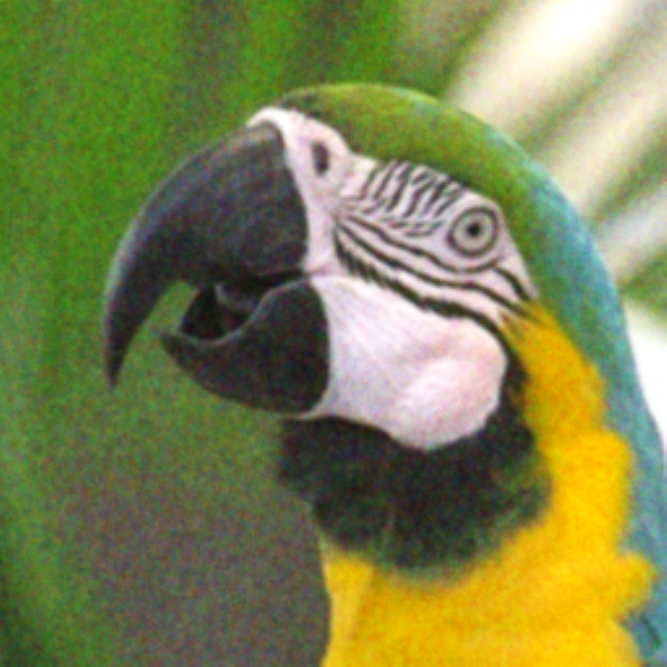}\hfill
    \includegraphics[width=0.39\linewidth]{Figures/JSD NSS/1.png}
    \caption{FG level = 0.100 ($1.03\mathrm{e}{-4}$)}
    \end{subfigure}  
    
\caption{Color film grain synthesis results on image "kodim23" from Kodak24 dataset. Ground truth grainy images on left, translated grainy images on right, with corresponding \acs{nss} histograms comparison. \acs{jsdnss} values between parenthesis.}
\label{fig:fg_levels}

\end{figure}

\section{Experimental results}
\label{sec:experimental}
In this section, first, the dataset used to train and evaluate our proposed models is presented, as well as the details of the training. Next, the film grain synthesis task is studied through quantitative and qualitative evaluations together with an ablation study. Then, the film grain removal task is explored in the same way, in addition to an evaluation of the filtering performance on real film grain.

\begin{figure*}
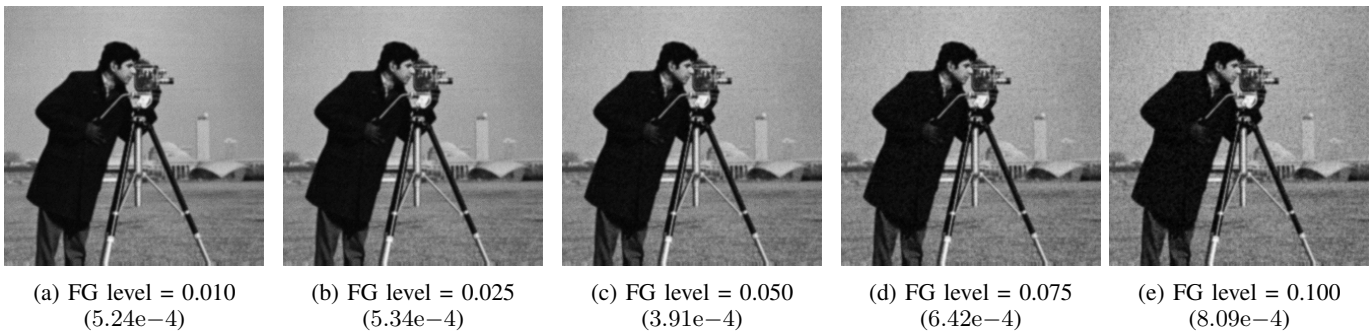

\centering
  \begin{subfigure}[b]{0.19\textwidth}
    \includegraphics[width=\textwidth]{Figures/Multi-level FG gray/01_fake_01.pdf}
    \caption{\centering FG level = 0.010 ($5.24\mathrm{e}{-4}$)}
    \label{fig:1}
  \end{subfigure}
  \hfill
  \begin{subfigure}[b]{0.19\textwidth}
    \includegraphics[width=\textwidth]{Figures/Multi-level FG gray/01_fake_025.pdf}
    \caption{\centering FG level = 0.025  ($5.34\mathrm{e}{-4}$)}
    \label{fig:1}
  \end{subfigure}
  \hfill
  \begin{subfigure}[b]{0.19\textwidth}
    \includegraphics[width=\textwidth]{Figures/Multi-level FG gray/01_fake_05.pdf}
    \caption{\centering FG level = 0.050  ($3.91\mathrm{e}{-4}$)}
    \label{fig:1}
  \end{subfigure}  
 \hfill
  \begin{subfigure}[b]{0.19\textwidth}
    \includegraphics[width=\textwidth]{Figures/Multi-level FG gray/01_fake_075.pdf}
    \caption{\centering FG level = 0.075 ($6.42\mathrm{e}{-4}$)}
    \label{fig:1}
  \end{subfigure}
  \begin{subfigure}[b]{0.19\textwidth}
    \includegraphics[width=\textwidth]{Figures/Multi-level FG gray/01_fake_1.pdf}
    \caption{\centering FG level = 0.100  ($8.09\mathrm{e}{-4}$)}
    \label{fig:1}
  \end{subfigure}
\caption{Gray film grain synthesis results at different intensity levels on image "01" from Set12 dataset. \acs{jsdnss} values between parenthesis.}
\label{fig:fg_gray}
\end{figure*}

\subsection{Experimental setting}
\subsubsection{Dataset construction}
\label{dataset}
To train our proposed models for film grain removal and synthesis, a large dataset of images was collected, including $400$ images from \ac{bsd} \cite{martin2001database}, $4744$ images from Waterloo Exploration Database \cite{ma2016waterloo}, $900$ images from DIV2K dataset \cite{Agustsson_2017_CVPR_Workshops}, $2650$ images from Flick2K dataset \cite{lim2017enhanced} and $140,000$ images from the \ac{kadis} \cite{kadid10k}. Therefore, we cover a large and diverse image space which enables the model to better generalize to unseen images. From each image, the maximum number of non-overlapping patches of size $256 \times 256$ is extracted. 

In order to have pairs of clean (grain-free) and grainy images to train our models, we used the publicly available code provided by Newson \textit{et al.} which consists in an implementation of the film grain rendering algorithm proposed in \cite{newson2017stochastic}, mentioned in Section~\ref{sec:rel}. To model film grain, authors employ an inhomogeneous Boolean model \cite{stoyan2013stochastic} that imitates the analog photographic process as closely as possible. The inhomogeneous boolean model corresponds to uniformly distributed disks using a Poisson process of variable intensity $\lambda$ which determines the amount of grain with respect to the local image gray level. To render film grain, they use a Monte Carlo simulation to determine the value of each output rendered pixel. The grain rendering algorithm is modeled with a single Gaussian kernel 
of variance $\sigma$ using a Monte Carlo simulation which performs simultaneous filtering and discretization of the film grain model. 

A wide range of grain types and intensities can be generated by varying the parameters of this model. The two main parameters are the average grain radius $\mu_{r}$ and its standard deviation $\sigma_{r}$. Some bigger values of these parameters accentuates the "grain" of the rendered result. The grainy image patches are obtained by adding film grain at five different intensities by varying the average grain radius $\mu_{r}$ in $\{0.010, 0.025, 0.050, 0.075, 0.100 \}$. 

To evaluate our film grain synthesis and filtering solutions on color images, we used the CBSD68 dataset composed of $68$ images of size $481 \times 324$ \cite{roth2009fields}, the Kodak24 dataset composed of $24$ color images of size $768 \times 512$ \cite{franzen1999kodak} and the McMaster dataset composed of $18$ images of size $500 \times 500$ \cite{zhang2011color}. For grayscale images we used the widely used Set12 dataset.

\subsubsection{Training parameters}
\hfill

Adam algorithm \cite{Adam1, Adam2} is used to train our models by optimizing the loss functions described in Eq. \eqref{eq:synthesis_loss} for film grain synthesis and Eq. \eqref{eq:filtering_loss} for film grain removal. The learning rate is fixed at $3\mathrm{e}{-4}$ for the U-Net with residual blocks for both tasks and $1\mathrm{e}{-4}$ for the PatchGAN discriminator. Batch size is equal to $1$ for film grain synthesis and $16$ for film grain removal.

\subsection{Film grain synthesis results}
\subsubsection{Quantitative and qualitative evaluations}
\hfill

For the film grain synthesis task, we have adopted the \ac{jsdnss} metric proposed by Li-Heng \textit{et al.} in \cite{chen2020learning} to evaluate our solution. The metric is based on \ac{nss} models in which, given a distorted and a clean image, \ac{mscn} coefficients are computed on local spatial neighborhoods of each image and their distributions are analyzed and compared. For natural images, such distributions behave normally, while distortions of different kinds perturb this regularity \cite{galdran2017retinal}.

Table \ref{tab:jsd_datasets} summarizes mean \ac{jsdnss} values computed between translated and corresponding ground truth grainy images on CBSD68, Kodak24 and McMaster datasets at each intensity level considered in the study, which were obtained by varying the average grain radius $\mu_{r}$. Mean \ac{jsdnss} values observed are very small, around $1\mathrm{e}{-3}$, which means that the compared distributions are close and therefore the compared images contain the same distortion type, i.e., film grain.

\begin{table}[]
\caption{Film grain synthesis results: mean \ac{jsdnss} results between ground truth and translated grainy images from CBSD68, Kodak24 and McMaster datasets.}
\centering
\begin{tabular}{l|ccccc}
\toprule
Dataset & \multicolumn{1}{c}{0.010} & \multicolumn{1}{c}{0.025} & \multicolumn{1}{c}{0.050} & \multicolumn{1}{c}{0.075} & \multicolumn{1}{c}{0.100} \\ \midrule
CBSD68           &0.0007                           &0.0007                           & 0.0008                           & 0.0009                           &0.0011                            \\ 
Kodak24          &0.0003                            &0.0003                            &0.0003                            & 0.0003                           &0.0004                            \\ 
McMaster         &0.0006                            &0.0007                            &0.0006                            &0.0004                            & 0.0004                           \\ \bottomrule
\end{tabular}
\label{tab:jsd_datasets}
\end{table}

Figure \ref{fig:fg_levels} visually compares color ground truth and translated grainy images at different intensity levels. Qualitative evaluation shows that the film grain map is not ignored but properly considered for controlling the intensity of the generated film grain. For each intensity level, \ac{nss} histograms are presented to compare distributions used for calculating the mean \ac{jsdnss} metric. Clean, ground truth and translated grainy images distributions are plot, from which we can observe that generated film grain distribution is closer and more similar to that of the ground truth grainy image than to the clean image one,  hence the small \ac{jsdnss} values obtained in Table \ref{tab:jsd_datasets}. Of course, generated film grain is not the exact same one as that of the ground truth, but they are hardly distinguishable perceptually. One can also see that distributions are wider for low intensity levels and narrower for higher film grain levels.

Considering the same model architecture and the same training details, a film grain synthesis model for grayscale images was trained where input channel is equal to 1 instead of 3 for color images. Figure \ref{fig:fg_gray} illustrates the resulting grainy images from our model at different intensity levels where from the lowest to the highest intensity level, film grain is more pronounced and accentuated.

\begin{table}[h!]
\centering
\caption{Mean \ac{jsdnss} comparison between ground truth and translated grainy images in terms of intensity levels on CBSD68, Kodak24 and McMaster datasets.}
\begin{tabular}{c|cccc}
\toprule
                                \multirow{2}{*}{Dataset}                                            &  \multirow{2}{*}{Generated FG level}         & \multicolumn{3}{|c}{Ground truth FG level}                                                    \\ \cmidrule{3-5}
         &  & \multicolumn{1}{|c}{0.010} & \multicolumn{1}{c}{0.050} & \multicolumn{1}{c}{0.100} \\ \midrule
\multicolumn{1}{c|}{\multirow{3}{*}{CBSD68}} &\multicolumn{1}{c|}{0.010}                         & \textbf{0.0007}            & 0.0008                     & 0.0014                     \\
                 \multicolumn{1}{c|}{}                  & \multicolumn{1}{c|}{0.050}                          & 0.0009                     & \textbf{0.0007}            & 0.0011                     \\
              \multicolumn{1}{c|}{}                  & \multicolumn{1}{c|}{0.100}                          & 0.0017                     & 0.0014                     & \textbf{0.0010}            \\ \midrule 
                \multicolumn{1}{c|}{\multirow{3}{*}{Kodak24}} &\multicolumn{1}{c|}{0.010}                          & \textbf{0.0003}            & 0.0004                     & 0.009                      \\
                  \multicolumn{1}{c|}{}                  &\multicolumn{1}{c|}{0.050}                         & 0.0004                     & \textbf{0.0003}            & 0.0005                     \\
                  \multicolumn{1}{c|}{}                  &\multicolumn{1}{c|}{0.100}                          & 0.0010                     & 0.0006                     & \textbf{0.0003}            \\ \midrule 
                  \multicolumn{1}{c|}{\multirow{3}{*}{McMaster}} & \multicolumn{1}{c|}{0.010}                         & \textbf{0.0006}            & 0.0007                     & 0.0014                     \\
                  \multicolumn{1}{c|}{}                  &\multicolumn{1}{c|}{0.050}                         & 0.0007                     & \textbf{0.0006}            & 0.0010                     \\
                 \multicolumn{1}{c|}{}                  &\multicolumn{1}{c|}{0.100}                         & 0.0013                     & 0.0008                     & \textbf{0.0004}\\   \bottomrule
\end{tabular}
\label{tab:jsd_levels}
\end{table}

To further investigate the model outputs in terms of intensity level, mean \ac{jsdnss} values are computed between ground truth grainy images and translated ones at three different intensity levels including $0.010$, $0.050$ and $0.100$, where the smallest \ac{jsdnss} values should be observed when ground truth and generated film grain levels match. Results are summarized in Table \ref{tab:jsd_levels}, where, in fact, the diagonal values are the smallest, confirming that distributions of generated and ground truth film grain at corresponding intensity level are the most similar. This demonstrates that our proposed model does not simply add some film grain but respects the film grain level specified at the input, based on which it controls the generated film grain intensity. In addition, the adopted metric is well suited to determining the presence, or absence of film grain, as well as to distinguishing between intensity levels.
\subsubsection{Ablation Study}
\hfill

For a more extensive analysis of our film grain synthesis solution, an ablation study was conducted to evaluate the contribution of the different components of the model including the role of the residual blocks in the generator and the role of the discriminator. To this end, three different configurations are evaluated and summarized in Table \ref{tab:fg_setting}. Config. 1: a basic U-Net architecture is optimized using an $\mathcal{L}_{L_{1}}$ loss. Config. 2: a \ac{cgan} based on a basic U-Net as generator and a PatchGAN as discriminator, optimized with a weighted sum of an $\mathcal{L}_{L_{1}}$ loss and an adversarial loss is used. Config. 3: a \ac{cgan} based on a U-Net with residual blocks as generator and a PatchGAN as discriminator, optimized with the same weighted sum as in Config. 2 is used. Config. 3 corresponds to our proposal.
\begin{table}[h!]
\centering
\caption{Settings of the three considered configurations in film grain synthesis ablation study.}
\label{tab:fg_setting}
\begin{tabular}{l|ccc}
\toprule
Settings & \multicolumn{1}{c}{Config. 1} & \multicolumn{1}{c}{Config. 2} & \multicolumn{1}{c}{Config. 3} \\ \midrule
U-Net  &  \Checkmark & \Checkmark & \Checkmark  \\
PatchGAN  & \XSolidBrush & \Checkmark & \Checkmark \\
Residual blocks  & \XSolidBrush &  \XSolidBrush  & \Checkmark \\ 
\bottomrule                         
\end{tabular}
\end{table}

\begin{figure}[t!]
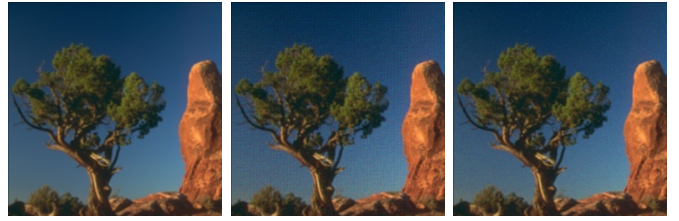

\centering
    \begin{subfigure}[]{0.32\linewidth}
        \includegraphics[width=\linewidth]{Figures/Ablation FG/unet.pdf}
        \caption{Config. 1}
    \end{subfigure}    
    \begin{subfigure}[]{0.32\linewidth}
        \includegraphics[width=\linewidth]{Figures/Ablation FG/pix2pix.pdf}
        \caption{Config. 2}
    \end{subfigure}
    \begin{subfigure}[]{0.32\linewidth}
        \includegraphics[width=\linewidth]{Figures/Ablation FG/proposed.pdf}
        \caption{Config. 3}
    \end{subfigure}
    \caption{Color film grain synthesis with the three different configurations considered in the ablation study (Table~\ref{tab:fg_setting}) on image "0058" from CBSD68 dataset.}
\label{fig:fg_abaltion}
\vspace{-2mm}
\end{figure}

Quantitative and qualitative evaluations of the three configurations considered in the ablation study are reported in Table \ref{tab:fg_abaltion} and Figure \ref{fig:fg_abaltion}. Table \ref{tab:fg_abaltion} reports mean \ac{jsdnss} values obtained with the different configurations. One can observe that the highest values are obtained with Config. 1, lower values by adding a discriminator in Config. 2, and much smaller values by adopting the residual blocks in the U-Net architecture in Config. 3. These values are well interpreted by the qualitative evaluation in Figure \ref{fig:fg_abaltion}, which shows the output of each configuration at intensity level $0.010$. 
Config. 1. produces blurred results and adds no film grain; hence the large \ac{jsdnss} values observed in Table \ref{tab:fg_abaltion} and that is mainly due to the exclusive use of the $\mathcal{L}_{L_{1}}$ loss in the training and optimization process. Config. 2 produces grainy images thanks to the discriminator but with an unpleasant appearance due to the lack of modeling capability. While Config. 3 produces realistic film grain thanks to the use of residual blocks. As it has already been proved theoretically and in various applications of ResNets, deeper and more efficient networks can be built with residual blocks resulting in better performance and rendering in film grain synthesis.
\vspace{-4mm}

\begin{figure*}[t!]
\centering
  \begin{subfigure}[b]{0.13\textwidth}
    \includegraphics[width=\textwidth]{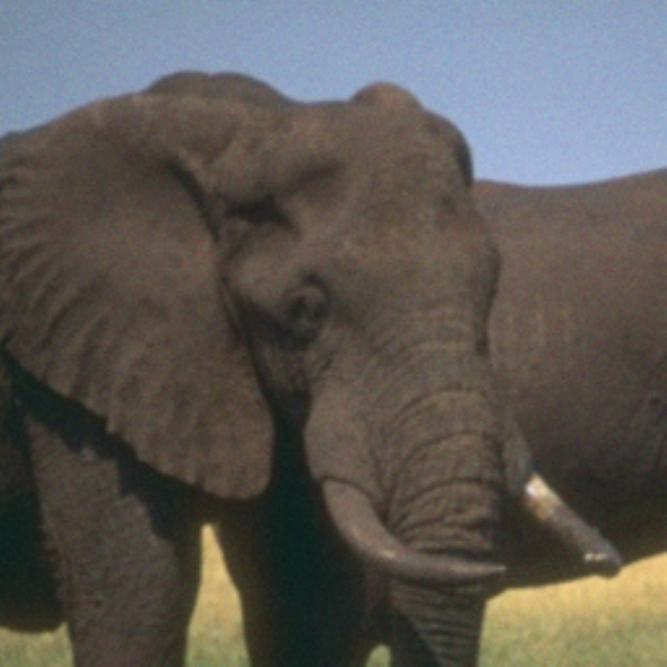}
    \caption{Grainy}
    \label{fig:1}
  \end{subfigure}
  \hfill
  \begin{subfigure}[b]{0.13\textwidth}
    \includegraphics[width=\textwidth]{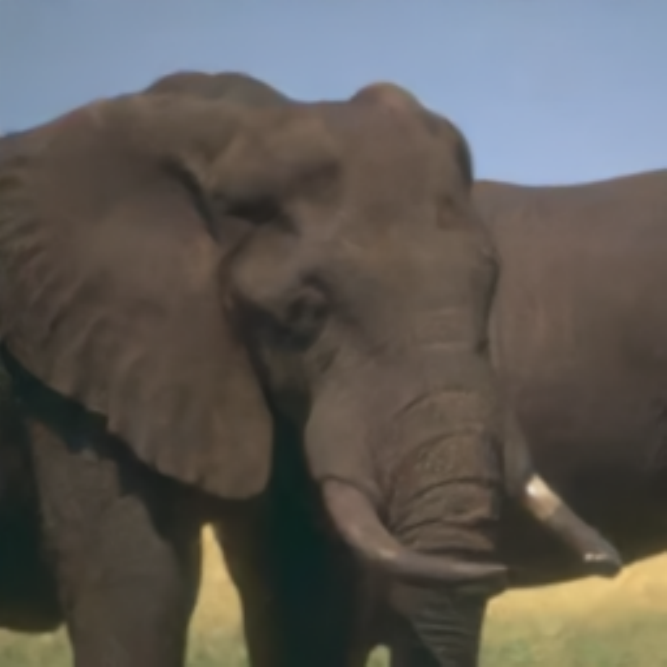}
    \caption{DnCNN}
    \label{fig:1}
  \end{subfigure}
  \hfill
  \begin{subfigure}[b]{0.13\textwidth}
    \includegraphics[width=\textwidth]{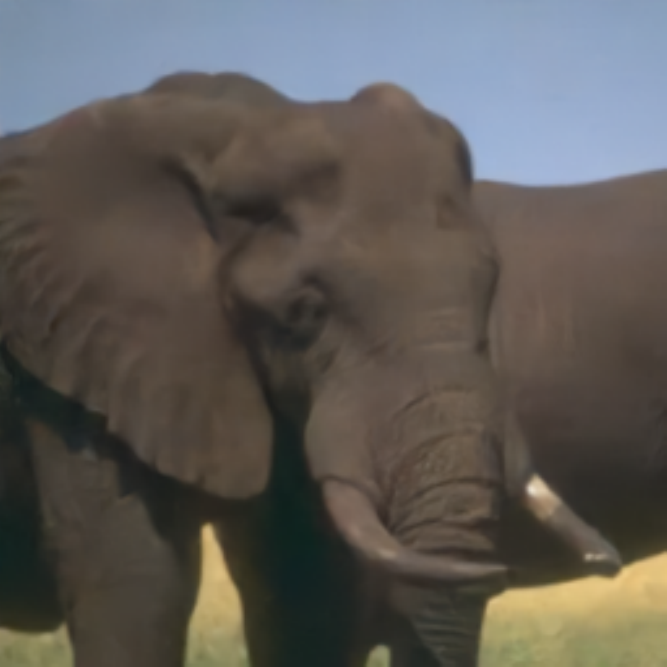}
    \caption{IRCNN}
    \label{fig:1}
  \end{subfigure}  
 \hfill
  \begin{subfigure}[b]{0.13\textwidth}
    \includegraphics[width=\textwidth]{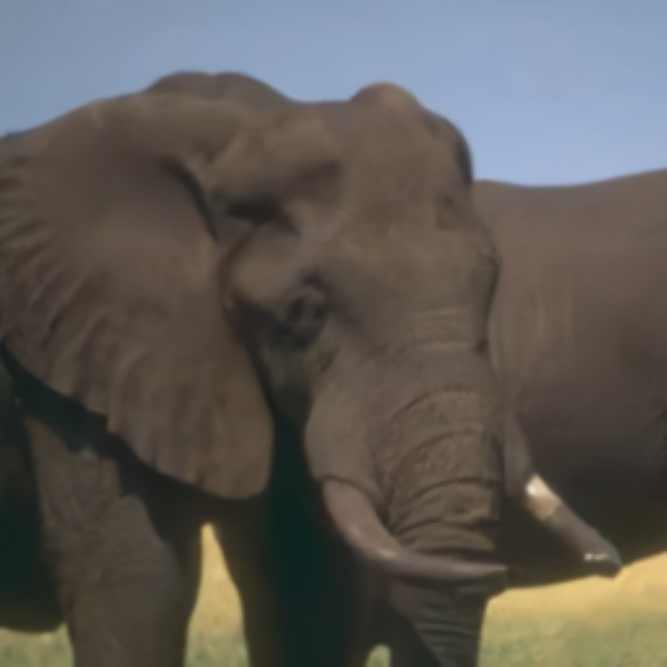}
    \caption{FFDNet}
    \label{fig:1}
  \end{subfigure}
  \hfill
  \begin{subfigure}[b]{0.13\textwidth}
    \includegraphics[width=\textwidth]{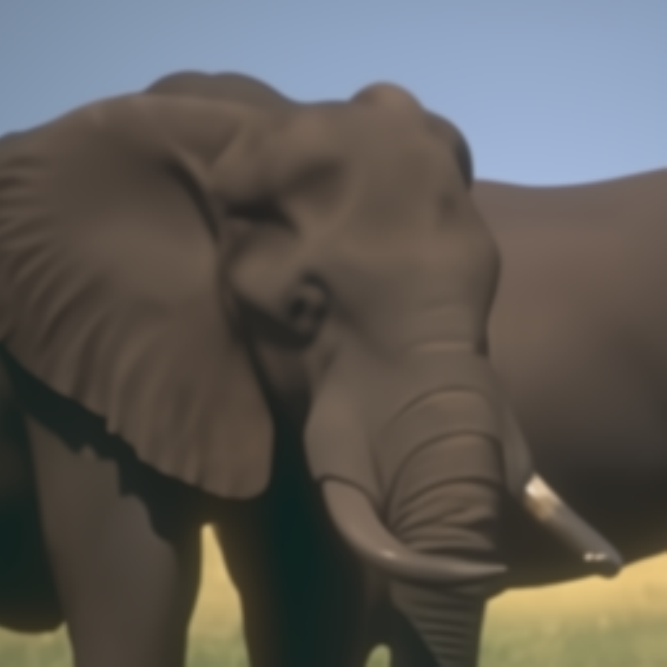}
    \caption{DRUNet}
    \label{fig:2}
  \end{subfigure}
    \hfill
  \begin{subfigure}[b]{0.13\textwidth}
    \includegraphics[width=\textwidth]{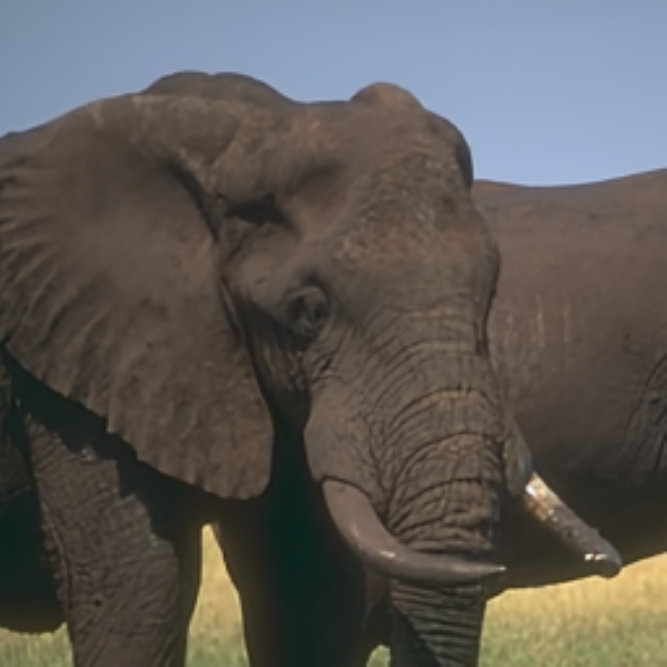}
    \caption{Ours}
    \label{fig:1}
  \end{subfigure}
  \hfill
  \begin{subfigure}[b]{0.13\textwidth}
    \includegraphics[width=\textwidth]{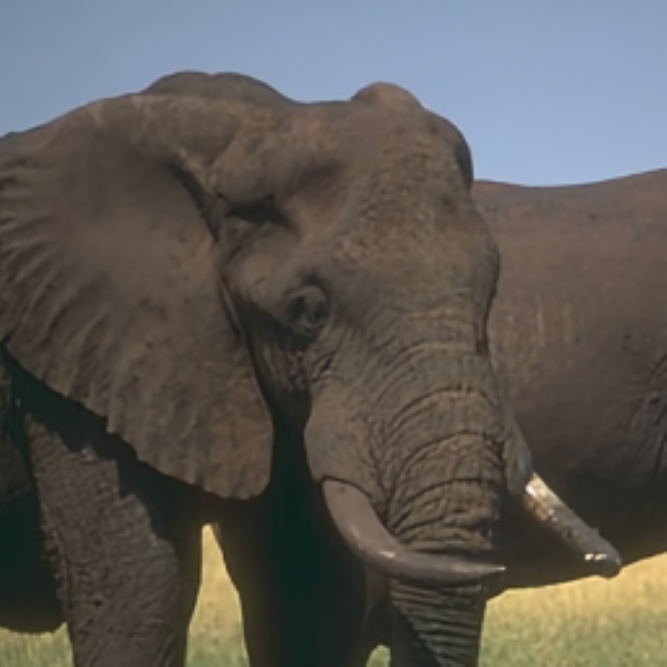}
    \caption{Ours (blind)}
    \label{fig:1}
  \end{subfigure}
    \vspace{.6ex}
  \begin{subfigure}[b]{0.13\textwidth}
    \includegraphics[width=\textwidth]{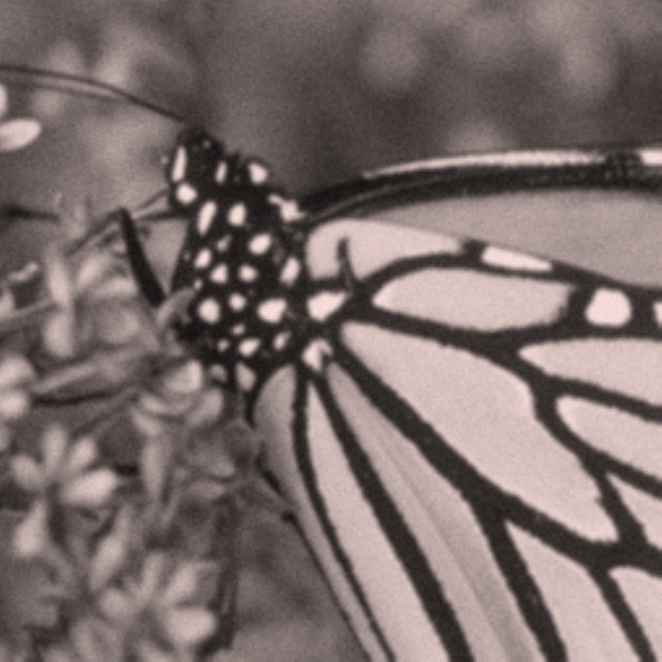}
    \caption{Grainy}
    \label{fig:1}
  \end{subfigure}
  \hfill
  \begin{subfigure}[b]{0.13\textwidth}
    \includegraphics[width=\textwidth]{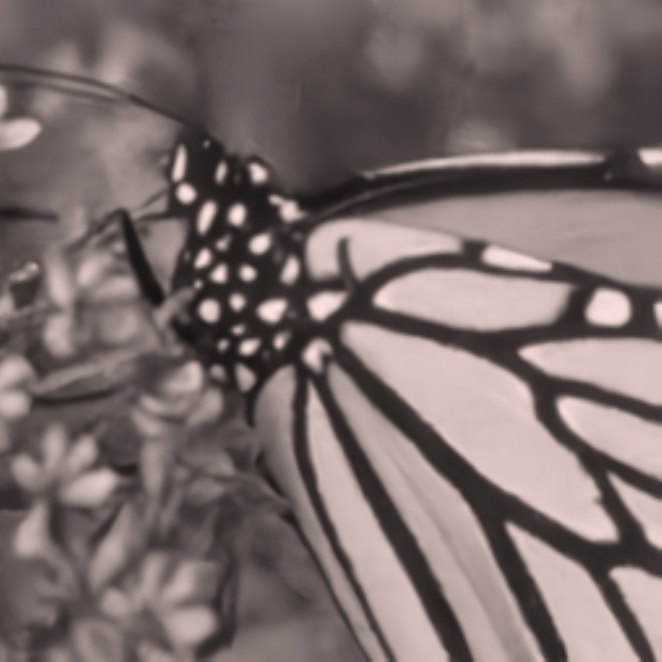}
    \caption{DnCNN}
    \label{fig:1}
  \end{subfigure}
  \hfill
  \begin{subfigure}[b]{0.13\textwidth}
    \includegraphics[width=\textwidth]{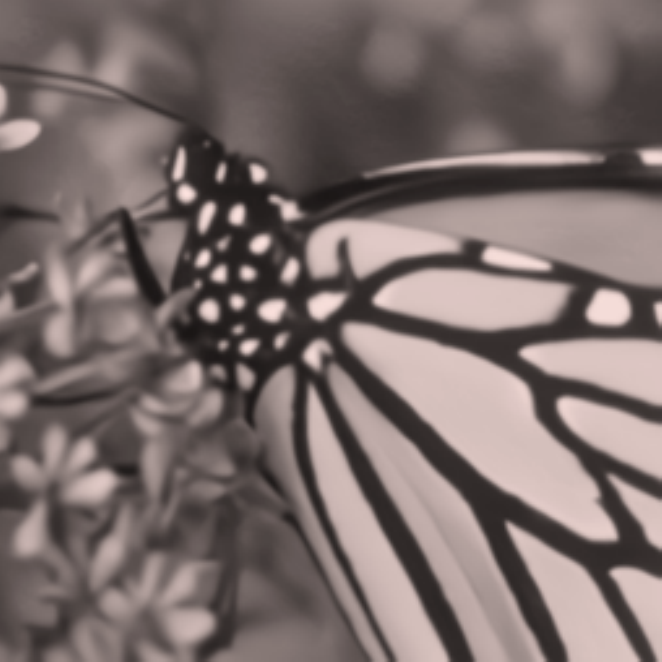}
    \caption{IRCNN}
    \label{fig:1}
  \end{subfigure}  
 \hfill
  \begin{subfigure}[b]{0.13\textwidth}
    \includegraphics[width=\textwidth]{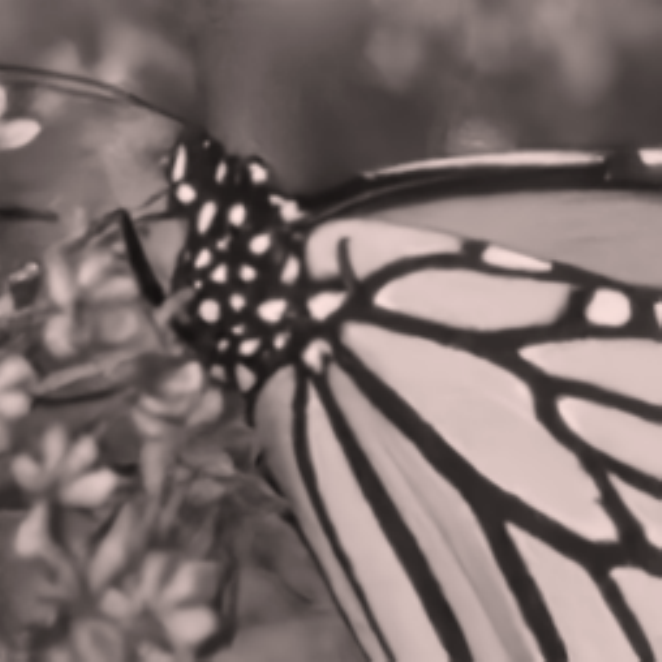}
    \caption{FFDNet}
    \label{fig:1}
  \end{subfigure}
  \hfill
  \begin{subfigure}[b]{0.13\textwidth}
    \includegraphics[width=\textwidth]{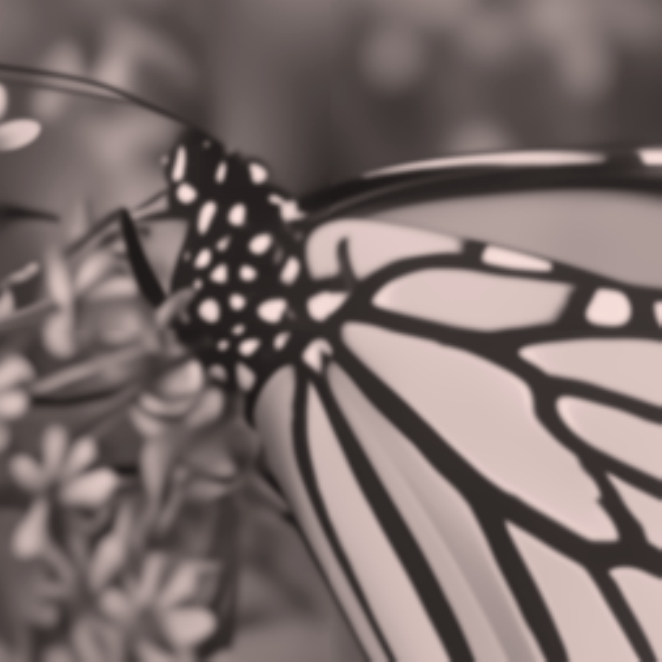}
    \caption{DRUNet}
    \label{fig:2}
  \end{subfigure}
    \hfill
  \begin{subfigure}[b]{0.13\textwidth}
    \includegraphics[width=\textwidth]{Figures/Gray comparison/ORG3.pdf}
    \caption{Ours}
    \label{fig:1}
  \end{subfigure}
  \hfill
  \begin{subfigure}[b]{0.13\textwidth}
    \includegraphics[width=\textwidth]{Figures/Gray comparison/ORG4.pdf}
    \caption{Ours (blind)}
    \label{fig:1}
  \end{subfigure}
\caption{Film grain removal results of different methods on image "0060" from CBSD68 dataset (top) and image "05" from Set12 dataset (bottom) with film grain level 0.010.}
\label{fig:comparison}
\end{figure*}
\subsection{Film grain removal results}
\subsubsection{Quantitative and Qualitative Evaluation}
\hfill

For the film grain removal task, we compared our proposed model with several state-of-the-art denoising methods, including one representative model-based method named \ac{bm3d} \cite{dabov2007image} for grayscale images and CBM3D \cite{dabov2007color} for color images, one CNN-based method which separately learns a single model for each noise level, i.e.,  DnCNN \cite{zhang2017beyond} and three CNN-based models that handle a wide range of noise levels, IRCNN \cite{zhang2017learning}, FFDNet \cite{zhang2018ffdnet} and DRUNet \cite{zhang2017learning}. 

Figure \ref{fig:comparison} visually illustrates the results for film grain removal on grayscale and color images. For color images, all state-of-the-art denoising methods are efficient in removing film grain even if they were designed a priori for Gaussian noise removal, because film grain and Gaussian noise are close both statistically and visually. Nevertheless, each of the methods introduces additional distortions while filtering film grain including color shift with IRCNN, blur with DnCNN and FFDNet, and tremendous loss of details and sharpness with DRUNet. As for the two versions of our proposed solution, they both recover most of the details with a slight color shift noticed with the blind configuration. For grayscale images, obviously, all methods successfully filter film grain. However, our proposed solutions provide much sharper outputs, with no perceptual difference. Moreover, in all configurations, no distortion can be noticed other than blurriness.

\begin{table}[h!]
\centering
\caption{Mean \ac{jsdnss} comparison between the three different configurations considered in the ablation study of film gain synthesis on CBSD68, Kodak24 and McMaster datasets.}
\label{tab:fg_abaltion}
\begin{tabular}{l|c|ccc}
\toprule
\multicolumn{1}{l|}{Dataset} & \multicolumn{1}{c|}{FG level} & \multicolumn{1}{c}{Config. 1} & \multicolumn{1}{c}{Config. 2} & \multicolumn{1}{c}{Config. 3} \\ \midrule
\multirow{5}{*}{CBSD68}  
                                 & 0.010                                  &0.0625                                      &0.0050                                     &\textbf{0.0007}                                     \\
                                      & 0.025                                  &0.0626                                     &0.0056                                     &\textbf{0.0007}                                       \\
                                      & 0.050                                  &0.0626                                     &0.0062                                      &\textbf{0.0008}                                       \\
                                      & 0.075                                  &0.0640                                     &0.0067                                      &\textbf{0.0009}                                       \\
                                      & 0.100                                  &0.0656                                      &0.0070                                      &\textbf{0.0011}                                      \\ \midrule
\multirow{5}{*}{Kodak24}               & 0.010                                  &0.0783                                      &0.0014                                        &\textbf{0.0003}                                      \\
                                      & 0.025                                  &0.0781                                     &0.0015                                      &\textbf{0.0003}                                      \\
                                      & 0.050                                  &0.0781                                      &0.0018                                      &0\textbf{.0003}                                      \\
                                      & 0.075                                  &0.0781                                      &0.0023                                      &\textbf{0.0003}                                      \\
                                      & 0.100                                  &0.0800                                       &0.0028                                      &\textbf{0.0004}                                      \\ \midrule
\multirow{5}{*}{McMaster}              & 0.010                                  &0.0590                                      &0.0041                                      &\textbf{0.0006}                                      \\
                                      & 0.025                                  &0.0598                                      &0.0046                                      &\textbf{0.0007}                                      \\
                                      & 0.050                                  &0.0614                                      &0.0056                                      &\textbf{0.0006}                                      \\
                                      & 0.075                                  &0.0660                                      &0.0056                                      &\textbf{0.0004}                                      \\
                                      & 0.100                                  &0.0695                                      &0.0058                                      &\textbf{0.0004}        \\ \bottomrule                         
\end{tabular}
\end{table}

Table \ref{tab:color_comparaison} reports the average \ac{psnr} and \ac{ssim} results of all state-of-the-art methods studied in this paper on CBSD68, Kodak24, McMaster and Set12 datasets. CNN-based methods have roughly equal performance with \ac{ffdnet} in the lead. The latter also yielded sharper results in the qualitative evaluation illustrated in Figure \ref{fig:comparison}. CBM3D, in contrast, provides higher \ac{psnr} and \ac{ssim} compared to deep learning models, simply because learning-based models tend to overfit the task they were trained to perform. Therefore, for a more accurate comparison, the best performing deep learning model for film grain removal was selcted, \ac{ffdnet}, and retrained to remove and filter film grain and named FFDNet - FG in Table \ref{tab:color_comparaison}. Better \ac{psnr} and \ac{ssim} results are observed with the fine-tuned model since it is evaluated on the same task it was trained for, but the performance is still not as efficient as our proposed solutions.

\begin{table}[h!]
\centering
\caption{Average \ac{psnr} (dB) / \ac{ssim} results of different film grain removal methods with an intensity level of 0.010 on CBSD68, Kodak24, McMaster and Set12 datasets.}
\begin{adjustbox}{max width=\linewidth}
\begin{tabular}{l | c c c | c }
\toprule
Dataset   & {CBSD68} & {Kodak24} & {McMaster} & Set12 \\  \midrule
CBM3D \cite{dabov2007color} $|$ BM3D \cite{dabov2007image}     &28.87 / 0.858                           &29.81 / 0.864                           &31.54 / 0.907                       &  29.27 / 0.854  \\  \midrule
DnCNN  \cite{zhang2017beyond}   & 27.74 / 0.787              & 28.72 / 0.801              & 29.76 / 0.844               & 28.00 / 0.820 \\  
IRCNN  \cite{zhang2017learning}   & 27.77 / 0.789              & 28.77 / 0.803               & 30.15 / 0.854         &  28.32 / 0.833 \\  
FFDNet \cite{zhang2018ffdnet}   & 27.80 / 0.792              & 28.83 / 0.807             & 30.22 / 0.858             &  28.01 / 0.822  \\
DRUNet  \cite{zhang2017learning}  & 27.70 / 0.779               & 28.73 / 0.779   & 30.25 / 0.855       &   27.97 / 0.812      \\ \midrule
FFDNet - FG  \cite{zhang2018ffdnet}  & 32.95 / 0.932               & 33.87 / 0.928   & 34.93 / 0.936    &-           \\ \midrule
Ours & \textbf{33.44 / 0.937} & \textbf{34.69 / 0.935}    & \textbf{36.04 / 0.948}     &     \textbf{33.37 / 0.926}       \\
Ours (blind)  &33.36 / 0.936 & 34.59 / 0.934    & 35.90 / 0.948       &  33.36 / 0.925         \\   \bottomrule 
\end{tabular}
\label{tab:color_comparaison}
\end{adjustbox}
\end{table}

Table \ref{tab:blind_nonblind} summarizes mean \ac{psnr} and \ac{ssim} results of our solutions (blind and non-blind) across all datasets and at all intensity levels considered. From these results, we can observe that performance of both models tends to deteriorate at higher intensity levels, which means that it is harder to restore details in the presence of strong film grain, as is the case with the state-of-the-art denoising models. Yet, \ac{psnr} and \ac{ssim} values are above 30 dB and around 0.900, respectively, which is quite acceptable. It can also be noticed that blind and non-blind models achieve almost similar performance, which implies that both learn to correctly map inputs to outputs. The slight difference in performance that can be observed is due to the additional information about intensity level provided to the non-blind configuration. 

\begin{figure*}[t]
\centering
\captionsetup{justification=centering}
    \begin{subfigure}[]{0.19\linewidth}
      \includegraphics[width=\linewidth]{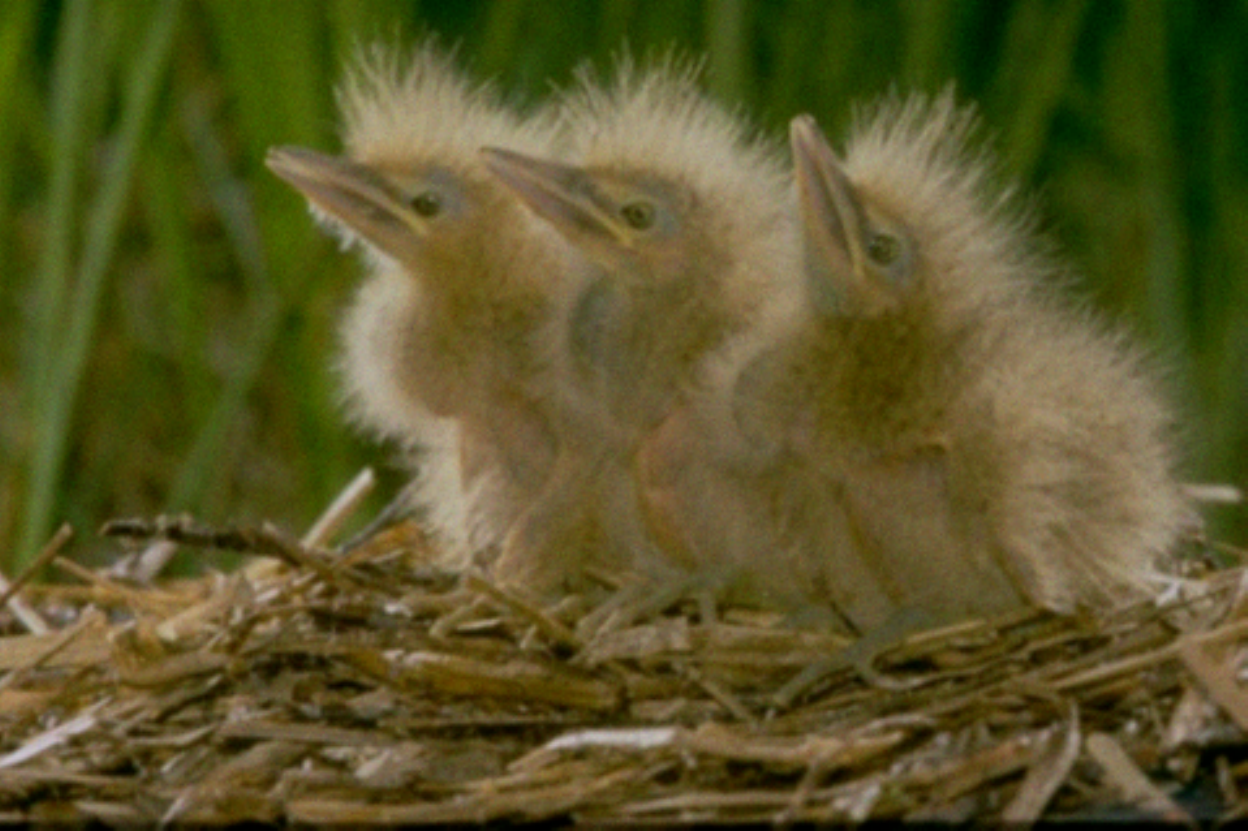}
    \caption{Grainy input \\ FG level = 0.010}
    \end{subfigure}\hfill
    \begin{subfigure}[]{0.19\linewidth}
      \includegraphics[width=\linewidth]{Figures/Blind_NonBlind_01/fake_ORG3_0.01_CBSD68_0031.pdf}
     \caption{Filtered (35.52dB)\\ ours (FG level = 0.010)}
     \label{fig:b}
    \end{subfigure}\hfill
    \begin{subfigure}[]{0.19\linewidth}
      \includegraphics[width=\linewidth]{Figures/Blind_NonBlind_01/fake_ORG3_0.05_CBSD68_0031.pdf}
     \caption{Filtered  (35.16dB)\\ ours (FG level = 0.050)}
     \label{fig:c}
    \end{subfigure}\hfill
    \begin{subfigure}[]{0.19\linewidth}
      \includegraphics[width=\linewidth]{Figures/Blind_NonBlind_01/fake_ORG3_0.1_CBSD68_0031.pdf}
     \caption{Filtered (33.95dB)\\ ours (FG level = 0.100)}
     \label{fig:d}
    \end{subfigure}\hfill
    \begin{subfigure}[]{0.19\linewidth}
      \includegraphics[width=\linewidth]{Figures/Blind_NonBlind_01/fake_ORG4_0.01_CBSD68_0031.pdf}
    \caption{Filtered (35.43dB)\\ ours (blind) }
     \label{fig:e}
    \end{subfigure}
    
    \vspace{.6ex}
    \begin{subfigure}[]{0.19\linewidth}
      \includegraphics[width=\linewidth]{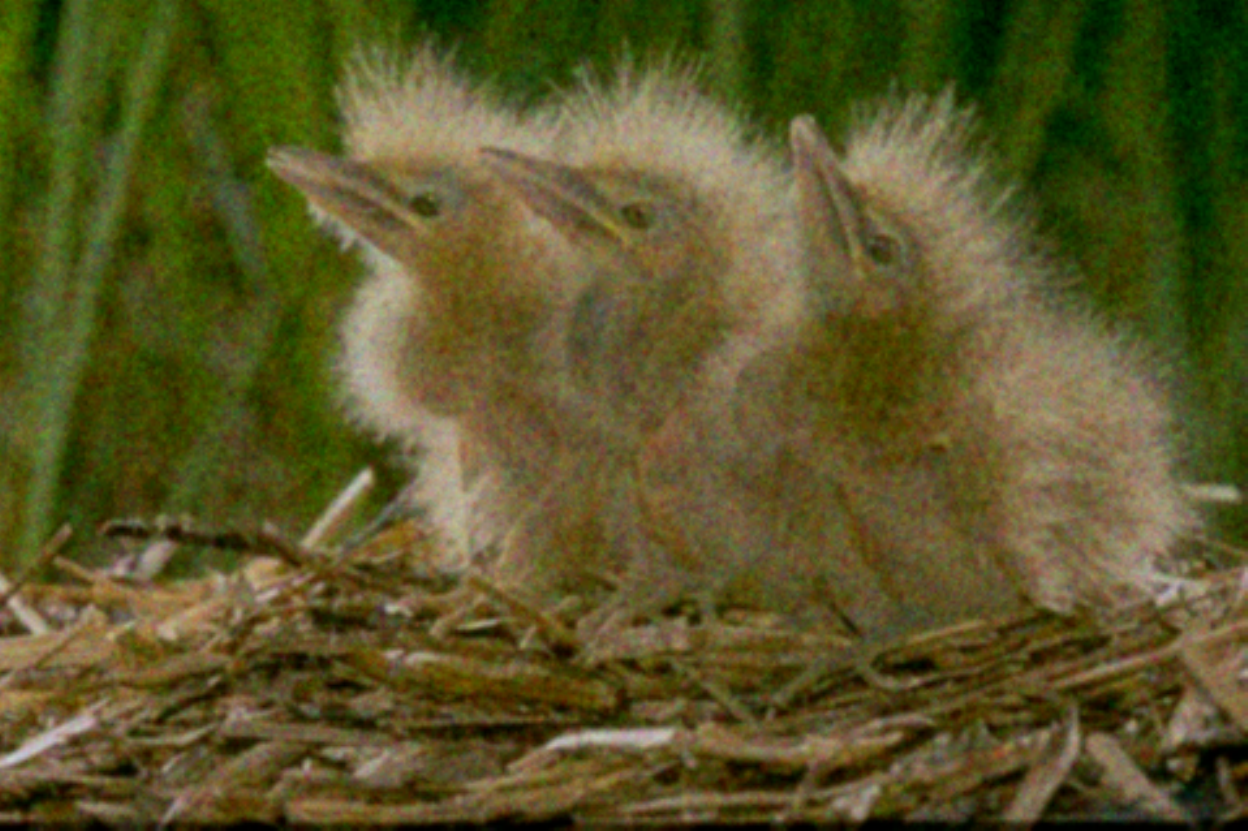}
    \caption{Grainy input \\ FG level = 0.100}
    \end{subfigure}\hfill
    \begin{subfigure}[]{0.19\linewidth}
      \includegraphics[width=\linewidth]{Figures/Blind_NonBlind_1/fake_ORG3_0.01_CBSD68_0031.pdf}
     \caption{Filtered (26.86dB)\\ ours (FG level = 0.010)}
         \label{fig:g}
    \end{subfigure}\hfill
    \begin{subfigure}[]{0.19\linewidth}
      \includegraphics[width=\linewidth]{Figures/Blind_NonBlind_1/fake_ORG3_0.05_CBSD68_0031.pdf}
     \caption{Filtered  (28.68dB)\\ ours (FG level = 0.050)}
     \label{fig:h}
    \end{subfigure}\hfill
    \begin{subfigure}[]{0.19\linewidth}
      \includegraphics[width=\linewidth]{Figures/Blind_NonBlind_1/fake_ORG3_0.1_CBSD68_0031.pdf}
     \caption{Filtered (32.53dB)\\ ours (FG level = 0.100)}
         \label{fig:i}
    \end{subfigure}\hfill
    \begin{subfigure}[]{0.19\linewidth}
      \includegraphics[width=\linewidth]{Figures/Blind_NonBlind_1/fake_ORG4_0.1_CBSD68_0031.pdf}
     \caption{Filtered (32.46dB)\\ ours (blind)}
          \label{fig:j}
    \end{subfigure}
    \caption{Color film grain removal using blind vs. non-blind versions of our proposed solution with variable film grain level maps on image "0031" from CBSD68 dataset: first row, input film grain level equals 0.01. Second row: input film grain level equals 0.1}
\label{fig:blindvsnonblind}
\end{figure*}

In order to investigate the relevance of the film grain level map in the non-blind configuration, the latter performance is evaluated by providing it with a film grain level map that is less than, equal to, and greater than the film grain level of a given grainy input image.
In Figure \ref{fig:blindvsnonblind}, a same grainy image but with different input grain level ($0.010$ for first row; $0.100$ for second row) is used as input to both the blind and non-blind models. For the latter, 3 different level maps are considered. In each row, we present the input grainy image (first column), 3 successive outputs of the non-blind model for different level maps (columns 2 to 4) and the output of the blind model (last column). In the first row, when the film grain level map matches the input image grain level, the best \ac{psnr} is obtained with a very well filtered output as shown in Fig.~\ref{fig:blindvsnonblind}(\subref{fig:b}). Whereas for a higher film grain level map, a lower \ac{psnr} is obtained with a tremendous loss of details and sharpness in the filtered outputs (see Fig.~\ref{fig:blindvsnonblind}(\subref{fig:c}) and (\subref{fig:d})). In the second row, when the film grain level map matches the input image grain level, the best \ac{psnr} is obtained along with the best perceptual quality as shown in Fig.~\ref{fig:blindvsnonblind}(\subref{fig:i}). On the other hand, for a lower film grain level map, a very low \ac{psnr} is obtained, and the model is not able to filter film grain properly (see Fig.~\ref{fig:blindvsnonblind}(\subref{fig:g}) and (\subref{fig:h})). This proves that the film grain level map is not ignored and is indeed considered by the non-blind model. On the other hand, the blind model performs well for both high and low film grain levels of grainy inputs without any information other than the grainy image itself.

\begin{table}[]
\centering
\caption{Average \ac{psnr} (dB) / \ac{ssim} results of our blind and non-blind models on CBSD68, Kodak24, McMaster and Set12 datasets.}
\label{tab:blind_nonblind}
\begin{adjustbox}{max width=\linewidth}
\begin{tabular}{c|c|c|c}
\toprule
Dataset                 & Level & Ours (non-blind) & Ours (blind) \\\midrule
\multirow{5}{*}{CBSD68} & 0.010 &33.44 / 0.937                 &33.36 / 0.936              \\ 
                        & 0.025 &33.39 / 0.934                  &33.31 / 0.934              \\  
                        & 0.050 &32.78 / 0.924                  &32.71 / 0.923             \\ 
                        & 0.075 &32.06 / 0.910                  &31.99 / 0.909              \\ 
                        & 0.100 &31.38 / 0.896                  &31.30 / 0.894              \\ \bottomrule 
\multirow{5}{*}{Kodak24} & 0.010 &34.69 / 0.935                &34.59 / 0.934              \\ 
                        & 0.025 &34.48 / 0.931                  &34.37 / 0.930             \\  
                        & 0.050 &33.90 / 0.922                  &33.79 / 0.920              \\ 
                        & 0.075 &33.17 / 0.908                 &33.06 / 0.906              \\ 
                        & 0.100 &32.45 / 0.894                  &32.35 / 0.892              \\ \bottomrule 
\multirow{5}{*}{McMaster} & 0.010 &36.04 / 0.948                 &35.90 / 0.948              \\ 
                        & 0.025 &35.70 / 0.945                  &35.61 / 0.945              \\  
                        & 0.050 &34.95 / 0.936                  &34.85 / 0.936              \\ 
                        & 0.075 &34.09 / 0.925                  &34.01 / 0.926              \\ 
                        & 0.100 &33.35 / 0.915                  &33.24 / 0.915              \\ \bottomrule 
\multirow{5}{*}{Set12} & 0.010 &33.37 / 0.926                 &33.36 / 0.925              \\ 
                        & 0.025 &33.06 / 0.920                  &33.04 / 0.920              \\  
                        & 0.050 &32.42 / 0.909                  &32.39 / 0.909              \\ 
                        & 0.075 &31.63 / 0.895                  &31.62 / 0.894              \\ 
                        & 0.100 &30.92 / 0.880                  &30.87 / 0.879              \\ \bottomrule 
\end{tabular}
\end{adjustbox}
\end{table}

\begin{table}[b!]
\centering
\caption{Settings of the three considered configurations in film grain removal ablation study.}
\label{tab:org_setting}
\begin{tabular}{l|ccc}
\toprule
Settings & \multicolumn{1}{c}{Config. 1} & \multicolumn{1}{c}{Config. 2} & \multicolumn{1}{c}{Config. 3}  \\ \midrule
U-Net  &  \Checkmark & \Checkmark & \Checkmark \\
Residual blocks  & \XSolidBrush &  \Checkmark  & \Checkmark  \\ 
$\mathcal{L}_{MS\textrm{-}SSIM} $ & \XSolidBrush  &\XSolidBrush & \Checkmark \\

\bottomrule      
\end{tabular}
\label{tab:four_settings}
\end{table}

\subsubsection{Ablation study}
\hfill

For further analysis of our film grain removal solution, an ablation study was conducted to evaluate the contribution of the different components of the network including the role of residual blocks and the mix loss function. Three different scenarii are investigated. In Config. 1: a basic U-Net architecture optimized with an $\mathcal{L}_{L_{1}}$ loss is proposed. In Config. 2: the same configuration is used, but with residual blocks in U-Net. Config. 3 corresponds to our proposed model: a U-Net with residual blocks optimized with a mix loss of $\mathcal{L}_{L_{1}}$ and $\mathcal{L}_{MS-SSIM}$ losses. Table~\ref{tab:four_settings} summarizes the settings of the three considered configurations. 

Quantitative evaluation of the three configurations considered on CBSD68 dataset is reported in Table \ref{tab:ablation_org}, from which we can observe that U-Net with residual blocks in Config. 2 achieves better performance than basic U-Net in Config. 1 thanks to the modeling capacity provided by residual blocks. Moreover, the model optimized using a mix loss function of $\mathcal{L}_{MS-SSIM}$ and $\mathcal{L}_{L_{1}}$ in Config. 3 achieves higher \ac{psnr} and \ac{ssim} than models optimized with only $\mathcal{L}_{L_{1}}$ in Config. 1 and 2. This is inline with the conclusion reached in \cite{7797130} which demonstrates that quality of the results can be improved significantly with better loss functions considering the same network architecture. Moreover, combining \ac{msssim} and $\mathcal{L}_{L_{1}}$ improved not only \ac{ssim} but also \ac{psnr}.
\vspace{-4mm}
\subsection{Generalization to unseen film grain}
\begin{figure*}
    \centering
    \begin{minipage}{.45\textwidth}
            \begin{subfigure}[t]{\textwidth}
                \includegraphics[width=\textwidth]{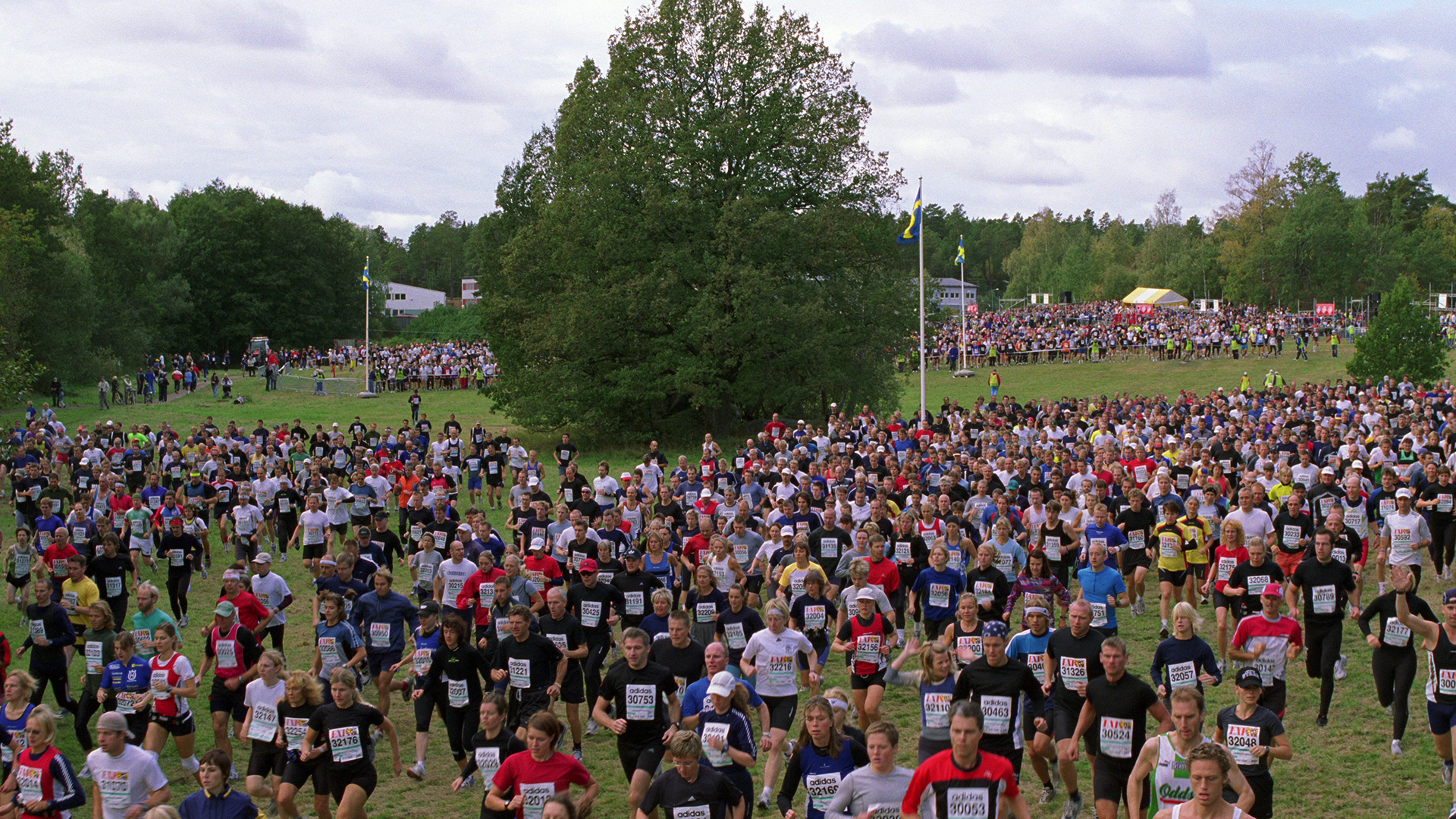}
                \label{fig:weather_activity}
            \end{subfigure}
        \end{minipage}
    \begin{minipage}{.13\textwidth}
        \begin{subfigure}[t]{\textwidth}
            \includegraphics[width=\textwidth]{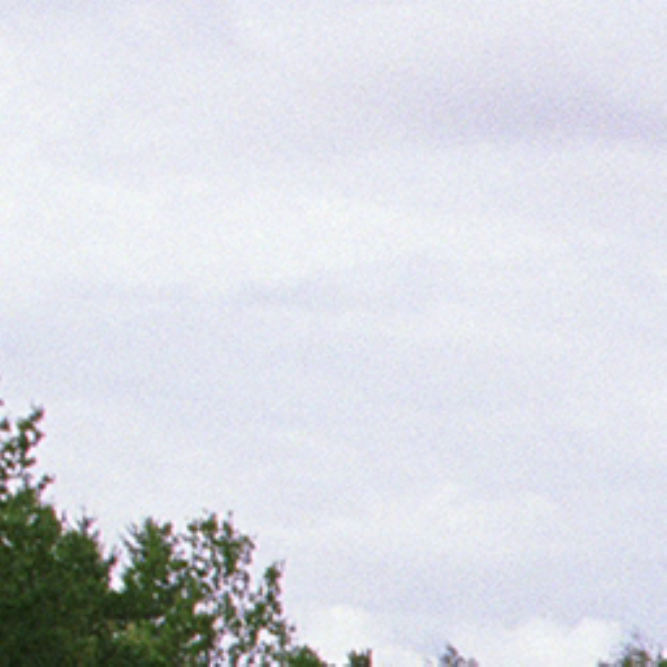}
            \label{fig:weather_filter1}
        \end{subfigure} \\
        \begin{subfigure}[b]{\textwidth}
            \includegraphics[width=\textwidth]{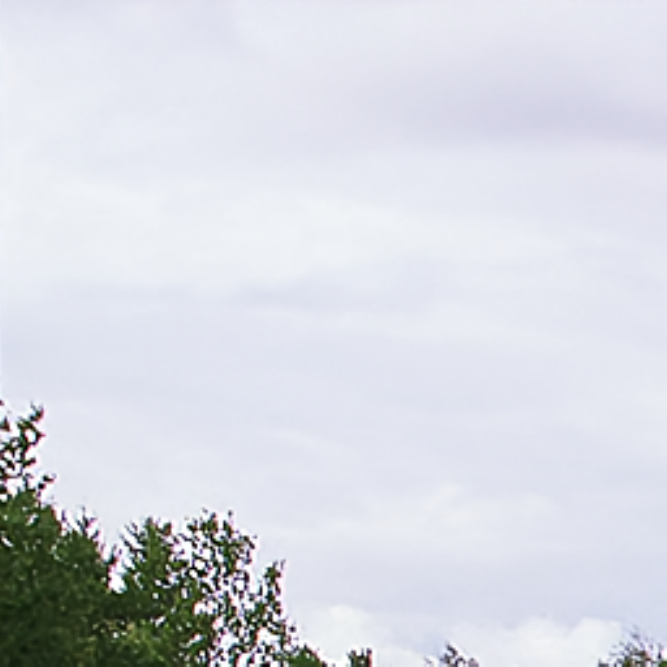}
            \label{fig:weather_filter2}
        \end{subfigure} 
    \end{minipage}
    \begin{minipage}{.13\textwidth}
        \begin{subfigure}[t]{\textwidth}
            \includegraphics[width=\textwidth]{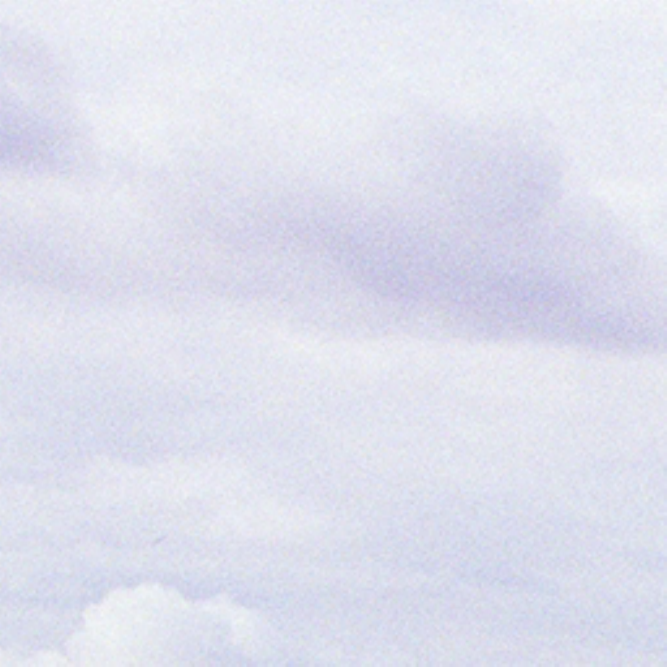}
            \label{fig:weather_filter1}
        \end{subfigure} \\
        \begin{subfigure}[b]{\textwidth}
            \includegraphics[width=\textwidth]{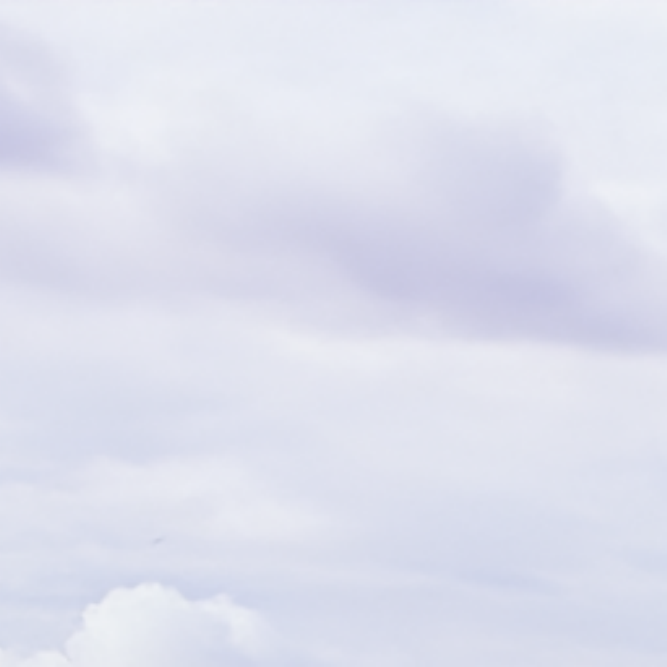}
            \label{fig:weather_filter2}
        \end{subfigure} 
    \end{minipage}
    \begin{minipage}{.13\textwidth}
        \begin{subfigure}[t]{\textwidth}
            \includegraphics[width=\textwidth]{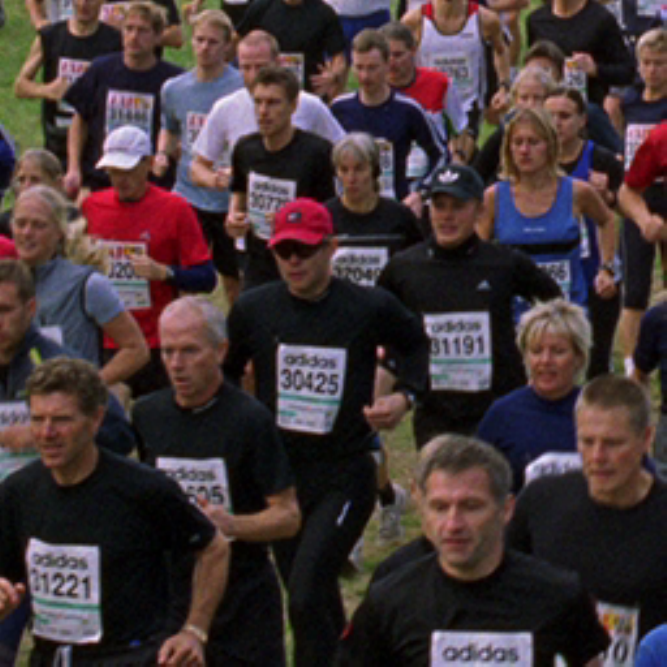}
            \label{fig:weather_filter1}
        \end{subfigure} \\
        \begin{subfigure}[b]{\textwidth}
            \includegraphics[width=\textwidth]{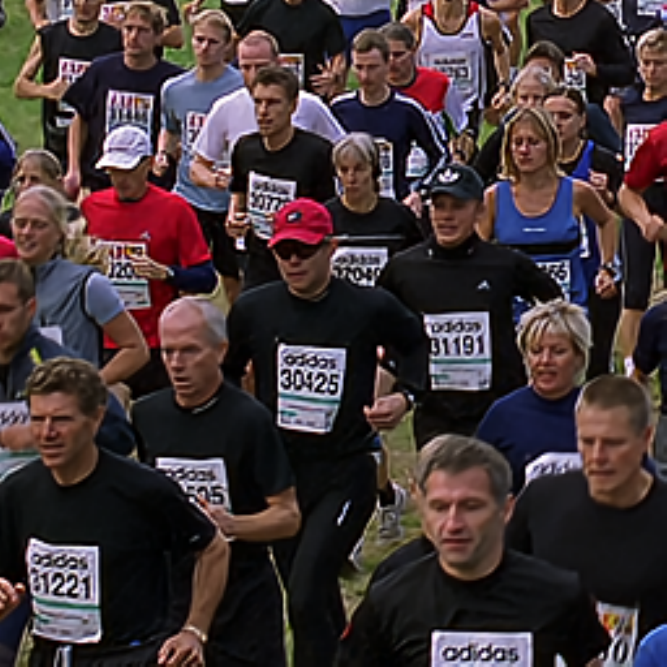}
            \label{fig:weather_filter2}
        \end{subfigure} 
    \end{minipage}
    \begin{minipage}{.13\textwidth}
        \begin{subfigure}[t]{\textwidth}
            \includegraphics[width=\textwidth]{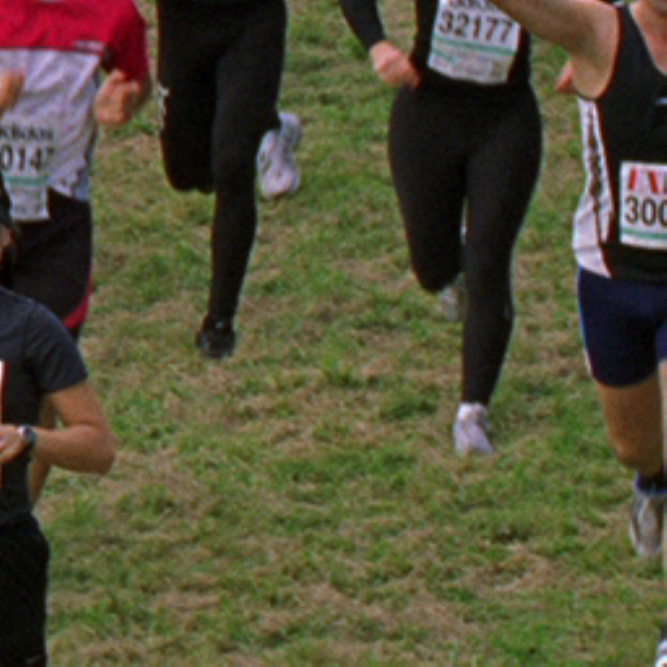}
            \label{fig:weather_filter1}
        \end{subfigure} \\
        \begin{subfigure}[b]{\textwidth}
            \includegraphics[width=\textwidth]{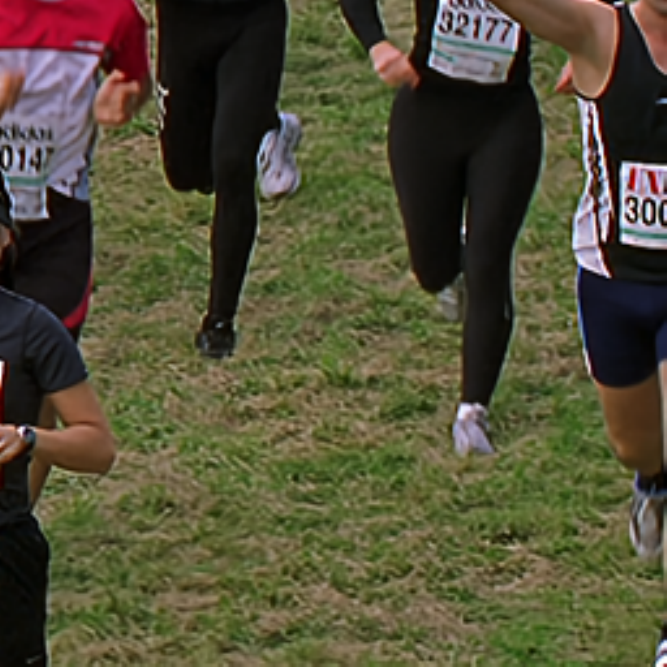}
            \label{fig:weather_filter2}
        \end{subfigure} 
    \end{minipage}
    \caption{Visual results of the blind film grain removal model on unseen film grain from CrowdRun sequence. On the left, the original frame with film grain and on the right, the cropped grainy patches (top) and cropped filtered patches (bottom).}
\label{fg:real_fg}
\end{figure*}

As our film grain removal model was trained on synthetic film grain generated using the same model, it is very important to test its limits on film grain coming from different sources. To do so, a frame from the CrowdRun test sequence that already contains film grain was selected to be filtered using the blind version of our solution. Figure \ref{fg:real_fg} shows a frame from the sequence and some grainy cropped patches and their corresponding filtered versions output by our blind model. One can see that film grain is properly filtered with no noticed loss in details or sharpness or blurriness. This test underlines the importance of developing a blind model that allows to filter film grain coming from different sources where the scale used to measure the level is probably not the same or simply unknown. 
\begin{table}[h!]
\centering
\caption{Average \ac{psnr} (dB) / \ac{ssim} results comparison with the three different configurations considered in the ablation study of film grain removal task on CBSD68 dataset.}
\begin{adjustbox}{max width=\linewidth}
\begin{tabular}{c|c|ccc}
\toprule
                & FG level & \multicolumn{1}{c}{Config. 1} & \multicolumn{1}{c}{Config. 2} & \multicolumn{1}{c}{Config. 3}  \\  \midrule
 \parbox[t]{2mm}{\multirow{5}{*}{\rotatebox[origin=c]{90}{CBSD68}}}                  
                                & 0.010                              &33.19 / 0.933               &33.42 / 0.935                                           &\textbf{33.44 / 0.937}                                                                                                        \\ 
                                  & 0.025                              &33.15 / 0.930             &33.37 / 0.933                                            &\textbf{33.39 / 0.934}                                                                                                     \\ 
                                  & 0.050                              &32.55 / 0.919               &32.75 / 0.922                                           &\textbf{32.78 / 0.924}                                                                                                      \\ 
                                  & 0.075                              &31.83 / 0.904              &32.01 / 0.907                                            &\textbf{32.06 / 0.910}                                                                                                      \\
                                  & 0.100                              &31.14 / 0.888               &31.32 / 0.892                                            &\textbf{31.38 / 0.896}                                                                                                      \\
                                 \bottomrule
\end{tabular}
\label{tab:ablation_org}
\end{adjustbox}
\end{table}
\vspace{-2mm}
\section{Conclusion}
\label{sec:conclusion}
In this paper, we are the first to propose to solve the film grain coding problem using deep learning. We have trained flexible and efficient deep learning models for film grain removal and synthesis tasks. We have shown that an encoder-decoder architecture is effective in removing film grain and provides better performance in terms of quality metrics when optimized using a combination of a pixel-to-pixel loss with a perceptual loss. As for film grain synthesis, \acp{gan} have proved to be effective in generating and synthesizing film grain with same statistics and similar look as the one in the training set. Controllable film grain intensity generation is supported by a conditional \ac{gan}, conditioned both on the input image and the intensity level, whereas flexible film grain removal is supported by a blind and a non-blind encoder-decoder architectures with competitive results. As future work, we seek to implement our models in an end-to-end video compression chain to measure bit-savings when using our film grain removal model and conduct subjective tests to evaluate the similarity between genuine and synthesized film grain.

\vspace{-4mm}
\bibliographystyle{ieeetr}
\bibliography{bibliography.bib}

\end{document}